\documentclass[review]{elsarticle}

\usepackage{lineno,hyperref}

\usepackage[english]{babel}
\usepackage[utf8]{inputenc}
\usepackage{amsmath}
\usepackage{graphicx}
\usepackage[colorinlistoftodos]{todonotes}
\usepackage{algorithm,algorithmic}
\usepackage{multirow}
\usepackage[toc,page]{appendix}

\modulolinenumbers[5]

\journal{Journal of \LaTeX\ Templates}

\begin{document}

\begin{frontmatter}

\title{Multivariate Density Estimation with Missing Data}

\author{Zhen Li, Lili Wu, Weilian Zhou and Sujit Ghosh}
\address{Department of Statistics, NC State University}



\address[mymainaddress]{2311 K. Stinson Drive, Raleigh, NC 27695-8203, USA}

\begin{abstract}
Multivariate density estimation is a popular technique in statistics with wide applications including regression models allowing for heteroskedasticity in conditional variances. The estimation problems become more challenging when observations are missing in one or more variables of the multivariate vector. A flexible class of mixture of tensor products of kernel densities is proposed which allows for easy implementation of imputation methods using Gibbs sampling and shown to have superior performance compared to some of the exisiting imputation methods currently available in literature. Numerical illustrations are provided using several simulated data scenarios and applications to couple of case studies are also presented.
\end{abstract}

\begin{keyword}
\texttt Gibbs Sampler\sep  Density Estimation\sep  Data Imputation\sep  Mixture Normal Models\sep  Conditional Density Function
\MSC[2010] 00-01\sep  99-00
\end{keyword}

\end{frontmatter}

\section{Introduction}
\label{sec:introduction}

Multivariate density estimation is one of the fundamental methods in statistics and has a long history when all observations are available to users. There are various density estimation techniques and methods  illustrated by Scott \cite{scott2015multivariate}. However, only limited methods are known when the analysts face with missing observations. To keep the exposition simple, we begin with the bivariate case. Suppose $(X_i, Y_i)\overset{\text{iid}}{\sim} f(x,y)$ for $i=1,...,n$, where $f(x, y)$ is a joint density function that needs to be estimated. We consider the case when some of the $X_i$'s or $Y_i$'s in the pair $(X_i, Y_i)$ are possibly missing and we assume that such observations are missing at random (MAR) according to the standard definition given by Rubin (2002) \cite{little2002statistical}. Models for uncertain data distributions based on mixture components through Bayesian approaches have been well studied for a long time. Some authors like Ferguson (1973, 1983) \cite{Ferguson}, Escobar \& West (1995) \cite{West} applied Bayesian methods using mixtures of Dirichlet processes and provided theoretical bases in univariate situation. Muller, Erkanli \& West (1996) \cite{Muller} generalized that work to multivariate framework, and used Gibbs sampler to do density estimation. 

However, those methods do not consider the situation where missing data exists. Therefore, we are motivated to estimate the density function using missing data and implement missing data imputation simultaneously. For $p\; (p\geq 1)$ dimensional data $\mathbf{x}=(\mathbf{x}_{\text{miss}},\mathbf{x}_{\text{obs}})$, we are in the framework of the mixture normal model $f_m(\boldsymbol{x}|\boldsymbol{\theta}, \boldsymbol{\lambda})=\sum_{k=1}^{m}\theta_k\prod_{i=1}^p\frac{1}{\lambda_{i}}\phi(\frac{x_i-s_{ik}}{\lambda_{i}})$, where $\boldsymbol{\theta}=(\theta_1,...,\theta_m)$, $\boldsymbol{\lambda}=(\lambda_1,...,\lambda_p)$, $\phi(\cdot)$ is the probability density function of a standard normal distributed random variable; $s_{ik}$ are suitably selected knots; the number of knots, $m$, is chosen by cross validation with mean square error criterion. We use prior on the parameters $\boldsymbol{\theta}\sim Dirichlet(\alpha_1, ..., \alpha_m)$ and $\lambda_i^2 \sim InverseGamma(a_i,b_i),\;i=1,...,p$. So we can sample missing values $\mathbf{x}_{\text{miss}}$ from the conditional distribution which can be derived from the above mixture normal model. With the spirit of Gibbs sampling, we can keep updating missing values and parameters $\boldsymbol{\theta}$ and $\boldsymbol{\lambda}$ so as to estimate the density function and then use conditional expectation to estimate the missing values. We will show more details of our method in Section 2.

In addition to employing those two Bayesian methods to impute the missing values, there are a lot of ways to do missing data imputation. Thus, we also compare our method with some other popular missing data imputation methods implemented in R packages such as ``predictive mean matching'' based on Buuren (2012, p. 73)\cite{Buuren}, which was proposed by Little (1988)\cite{Little}. The first one uses predictive mean matching which uses linear regression to get predictions on the missing values for numeric variables, which is implemented by one of the commonly used R package ``mice" (Multivariate Imputation via Chained Equations)\cite{mice}. The second way is to use function ``aregImpute" by default in the R package ``Hmisc"\cite{Hmisc}, which uses predictive mean matching with optional weighted probability sampling of donors rather than using only the closest match. The third method is a nonparametric missing value imputation using random forest which is implemented by the R package "missForest"\cite{missForest}. The fourth method is from the R package ``Amelia"\cite{Amelia}, which runs the bootstrap EM algorithm on incomplete data and creates imputed datasets. The last one is using the R package ``mi"\cite{mi} (Multiple imputation with diagnostics) which builds multiple imputation models to approximate missing values and also uses predictive mean matching method. The above five R packages are the most widely used tools for the missing data imputation nowadays and we will compare ours with them through some simulation results for two dimensional data and real data application for four dimensional cases in section 3. And finally we will discuss some future work in section 4.

\section{Methodology}
\label{sec:theory}
\subsection{Problem Setup}
Suppose we have data $\mathbf{X}=(X_1,X_2,...,X_p)$ and we want to use $f(X_1,X_2,...,X_p|\boldsymbol{\theta})$ to estimate the joint density of $\mathbf{X}$. If our data is complete, there are several ways to deal with this problem. But if part of our data is missing, those methods are not available. So our method is developed under the motivation to solve the problem of density estimation with missing data issue.

\subsection{Bayesian Mixture Density Estimation}
Ferguson (1973, 1983) \cite{Ferguson}, Escobar \& West (1995) \cite{West} and Erkanli \& West (1996) \cite{Muller} applied Bayesian methods using mixtures of Dirichlet processes to do density estimation. For $p$ dimensional data $\boldsymbol{x}=(x_1,...,x_p)'$, $f_m(\boldsymbol{x}|\boldsymbol{\theta},\boldsymbol{
\lambda})=\sum_{k=1}^{m}\theta_k\prod_{i=1}^p\frac{1}{\lambda_{i}}\phi(\frac{x_i-s_{ik}}{\lambda_{i}})$ can be used to approximate $f(\boldsymbol{x})$ typically, where $\phi(\cdot)$ is the probability density function of a standard normal distributed random variable; m is the number of modes; $s_{ik}$ are suitably selected knots from the data (see details in section 2.3), $i=1,...,p$, $k=1,...,m$. 

In order to estimate $f(\boldsymbol{x})$, we can first choose $m$ and $s_{ik}$ suitably and use the Bayesian method to estimate $\boldsymbol{\theta}$ and $\boldsymbol{\lambda}$. Typical priors for $\boldsymbol{\theta}$ and $\boldsymbol{\lambda}$ are $\boldsymbol{\theta} \sim Dir(\alpha_1,..., \alpha_m)$ and $\lambda_i^2 \sim InverseGamma(a_i,b_i),\;i=1,...,p$; $\boldsymbol{\theta}, \lambda_1,...,\lambda_p$ are independent. Then, we can use Gibbs sampler to sample $\boldsymbol{\theta}$ and $\boldsymbol{\lambda}$ given the data to estimate the density. The sampling method is shown in Algorithm \ref{alg:TBMDE}. To implement Algorithm \ref{alg:TBMDE}, we need Algorithm \ref{alg:condTheta} and Algorithm \ref{alg:condLambda} for conditional posterior sampling of $\boldsymbol{\theta}$ and $\boldsymbol{\lambda}$ which are in the Appendix.

\begin{algorithm}[H]
        \begin{algorithmic}[1]
            \STATE Suppose we have $p$ dimensional complete data $\mathbf{x}_1,...\mathbf{x}_n$, $f(\mathbf{x}|\boldsymbol{\theta},\boldsymbol{\lambda})=\sum_{k=1}^m \theta_k f_k(\mathbf{x}|\boldsymbol{\lambda})$.\\ $\boldsymbol{\theta} \sim Dir(\alpha_1,..., \alpha_m)$, $\lambda_i^2 \sim InverseGamma(a_i,b_i),\;i=1,...,p$ and $\boldsymbol{\theta}$, $\{\lambda_i\}_{1\leq i \leq p}$ are independent.
            \STATE Initialize $\boldsymbol{\theta}^{(0)}=(\frac{1}{m},\frac{1}{m},...,\frac{1}{m})$ and $\boldsymbol{\lambda}^{2(0)}$. 
            \FOR {iteration $l=1,2,...$} 
            \STATE Sample $\boldsymbol{\theta}^{(l)} \sim \boldsymbol{\theta}|\boldsymbol{\lambda}^{(l-1)}$ using \textbf{Algorithm 3}.
            \STATE Sample $\boldsymbol{\lambda}^{2(l)} \sim \boldsymbol{\lambda}^2|\boldsymbol{\theta}^{(l)}$ using \textbf{Algorithm 4}.
            \ENDFOR
            \end{algorithmic}
            \caption{Typical Bayesian Mixture Density Estimation (TBMDE)}
            \label{alg:TBMDE}
            \end{algorithm}

\subsection{Gibbs Mixture Data Imputation (GMDI)}
In practice, observations are sometimes missing in one or more variables of the multivariate vectors. However, we hope to make use of the observed part of the missing data to estimate the density function and impute the missing data simultaneously. For this purpose, we cannot use TBMDE which is based only on complete data. In this section, we will go through Gibbs Mixture Data Imputation (GMDI) in details which guarantees us to make use of both the complete data and the observed variables of the missing data. 

First, we will show how we choose the parameters in the model. For the number of the knots $m$, we use cross validation to choose it, which will be shown in Section 2.4. For the knots $\{s_{ik}\}_{1\leq i\leq p,1\leq k\leq m}$, they are chosen as following:
\begin{itemize}

\item
Choose knots $s_{ik},i=1,...,p, k=1,...,m$: $\mathbf{s_{i}}=(s_{i1}, s_{i2},...,s_{im})'$ are the $m$ knots of variable $X_i$. We pick the knots for $X_1$ first, $\mathbf{s_{1}}=(s_{11}, s_{12},...,s_{1m})'$ is where $s_{11}=\text{min}(X_1),s_{1m}=\text{max}(X_1)$, and $s_{1j}=X_1^{(l_j)},j=2,...,m-1$, which is the index of the ordered value for variable $X_1$, $l_j=[\frac{j-1}{m-1}n]$,  and $n$ is the sample size. After determining the knots in $X_1$, we can set the values of the left variables in the same sample as corresponding knots, i.e. $(s_{1k},s_{2k},...,s_{pk})$ is a sample from the data, $k=1,...,m$.
\end{itemize}

From the multidimensional mixture normal model

\begin{equation}
f_m(\boldsymbol{x}|\boldsymbol{\theta},\boldsymbol{\lambda})=\sum_{k=1}^{m}\theta_k\prod_{i=1}^p\frac{1}{\lambda_{i}}\phi(\frac{x_i-s_{ik}}{\lambda_{i}}),
\end{equation}
we can derive the conditional density function. For a missing item of the data, 
\begin{equation}
p(x_{i,miss},\;i\in M|\boldsymbol{\theta},\boldsymbol{\lambda},x_{j,obs},\;j \notin M)=\sum_{k=1}^{m}\theta_k' \prod_{i\in M}\frac{1}{\lambda_i}\phi(\frac{x_{i,\;miss}-s_{ik}}{\lambda_i})
\end{equation}
where $\theta_k'=\frac{\theta_k \prod_{j\notin M}\phi(\frac{x_{j,\;obs}-s_{jk}}{\lambda_j})}{\sum_{k=1}^{m}\theta_k \prod_{j\notin M}\phi(\frac{x_{j,\;obs}-s_{jk}}{\lambda_j})}$, $1\leq k \leq m$, and $M$ is the missing value index set of $\boldsymbol{x}$, and $M\subset \{1,...,p\}$. 

Therefore, in each iteration, we can sample missing values given the observed values and current $(\boldsymbol{\theta},\boldsymbol{\lambda})$, and then we can sample a new $(\boldsymbol{\theta},\boldsymbol{\lambda})$ given the current sampled missing values and observed values. We keep updating the missing values and $(\boldsymbol{\theta},\boldsymbol{\lambda})$ in this way under the same spirit of the Gibbs sampler. Same as TBMDE, the prior for $\boldsymbol{\theta}$ is $\boldsymbol{\theta} \sim Dir(\alpha_1,...,\alpha_m)$ and the prior for $\boldsymbol{\lambda}$ is $\lambda_i^2 \sim InverseGamma(a_i,b_i),\;i=1,...,p$ and $\{\lambda_i\}_{1\leq i \leq p}$ are independent. Algorithm \ref{alg:GMDI} shows steps of our modified Gibbs sampler.

\begin{algorithm}[H]
        \begin{algorithmic}[1]
        \STATE Suppose we have $p$ dimensional complete or missing data $\mathbf{x}_1,...\mathbf{x}_n$.
            \STATE Initialize $\boldsymbol{\theta}^{(0)}=(\frac{1}{m},\frac{1}{m},...,\frac{1}{m})$ and $\boldsymbol{\lambda}^{2(0)}$.
            \FOR {iteration $l=1,2,...$} 
            \FOR {$i=1,...,n$}
            \FOR {$j\in M_i$ (Missing value index set of $\mathbf{x_i}$)}
            \STATE Sample $X_{ij,miss}^{(l)}\sim p(x_{ij,miss},\;j\in M_i|\boldsymbol{\theta}^{(l-1)}, x_{ij,obs},\;j \notin M_i)$.
            \ENDFOR
            \ENDFOR
            \STATE Sample $(\boldsymbol{\theta}^{(l)},\boldsymbol{\lambda}^{2(l)}) \sim (\boldsymbol{\theta},\boldsymbol{\lambda}^2)|\mathbf{X}_{obs}, \mathbf{X}_{miss}^{(l)}$ using \textbf{Algorithm 1}.
            
            \ENDFOR
        \end{algorithmic}
      \caption{Gibbs Mixture Data Imputation (GMDI)}
      \label{alg:GMDI}
\end{algorithm}

In our modified Gibbs sampler, we sample $(\boldsymbol{\theta},\boldsymbol{\lambda})$ based on observed values and missing values, so we can make use of the data information as much as we can. After obtaining a sampled $(\boldsymbol{\theta},\boldsymbol{\lambda})$, we can impute the missing value of $x_i$ in $\boldsymbol{x}$ using
\begin{equation}
E(X_{i,\;miss}|\boldsymbol{\theta},\boldsymbol{\lambda},X_{j,\;obs}=x_{j,\;obs},\;j\notin M)=\sum_{k=1}^m \theta_k' s_{ik},\;\; \text{for}\;\;\forall i\in M,
\end{equation}
where $M$ is the missing value index set of $\boldsymbol{x}$.

\subsection{Cross Validation}
When we set up the model, a natural question is how many knots we need to choose, i.e. $m$, in the mixture normal model. The cross validation is a widely used way to deal with this problem.
\subsubsection{Overview of Cross Validation}
Cross validation \cite{Jun} is a model validation technique for assessing how the results of a statistical analysis will generalize to an independent data set. It is mainly used in settings where the goal is prediction, and one wants to estimate how accurately a predictive model will perform in practice. In a prediction problem, a model is usually given a dataset of known data on which training is run (training dataset), and a dataset of unknown data (or first seen data) against which the model is tested (called the validation dataset or testing set). The goal of cross validation is to define a dataset to ``test" the model in the training phase (i.e., the validation set), in order to limit problems like overfitting, give an insight on how the model will generalize to an independent dataset (i.e., an unknown dataset), etc.
\subsubsection{Implement Cross Validation}
In our experiment, we use 5-fold cross validation to choose the best number of knots. We divide the data into 5 parts evenly and set one of them as the test data set $TE$ and the other 4 parts as the training data set $TR$. For a fixed number of knots $m$, we use the training data to produce a series of $(\boldsymbol{\theta^{(l)}},\boldsymbol{\lambda^{(l)}})$ using the above Algorithm, $1\leq l\leq L$, where $L$ is the number of posterior samples. For a sample $\boldsymbol{x}_t=(x_{t1},...,x_{tp})' \in TE$, $t=1,...,n$, we denote $C_t$ as the non-missing value index set of $\boldsymbol{x}_t$. Then for $i\in C_t$, we estimate $x_{ti}$ by
\begin{equation}
\begin{aligned}
\hat{x}_{ti}&=\frac{1}{L}\sum_{l=1}^L E(X_i|\boldsymbol{\theta^{(l)}},\boldsymbol{\lambda^{(l)}},X_j=x_{tj},j\in C_t \backslash \{i\})\\
&=\begin{cases} \frac{1}{L} \sum_{l=1}^L \sum_{k=1}^m \theta_k^{(l)} s_{ik},\;\;\;\text{if}\;C_t=\{i\},\\
\frac{1}{L} \sum_{l=1}^L \sum_{k=1}^m \frac{\theta_k^{(l)} \prod_{j\in C_t\backslash \{i\}} \phi(\frac{x_{tj}-s_{jk}}{\lambda_j^{(l)}})}{\sum_{k=1}^m \theta_k^{(l)} \prod_{j\in C_t\backslash \{i\}} \phi(\frac{x_{tj}-s_{jk}}{\lambda_j^{(l)}})} s_{ik},\;\;\;\text{if}\;C_t \neq \{i\}. \end{cases}
\end{aligned}
\end{equation}

Then we compute the scaled sum of squared of error of the test data set
\begin{equation}
\label{sSSE}
\text{sSSE}=\sum_{\boldsymbol{x_t}\in TE} \sum_{i\in C_t} \frac{(\hat{x}_{ti}-x_{ti})^2}{\Lambda_i},
\end{equation}
where $\Lambda_i$ is the sample variance of the observed values of the $i^{th}$ variable, $i=1,...,p$.
Since our method is 5-fold cross validation, for each test data set we can calculate the $\text{sSSE}_r, 1\leq r \leq 5$, using (\ref{sSSE}). Since $\sum_{r=1}^5 \text{sSSE}_r$ is a measurement for the behavior of the model when the number of knots is fixed, we calculate this measurement for different $m$ and choose the $m$ that has the minimal value of $\sum_{r=1}^5 \text{sSSE}_r$ in our experiment.

\section{Results and Analysis}
\subsection{Data Generation}
We use three types of data sets to test the effectiveness of GMDI. They are from simulation, the real data set ``airquality" in R package ``mice" and the real data set ``Iris'' \cite{Iris}.
\subsubsection{Simulation Data Set-up}
\begin{figure}
\centering
\includegraphics[width=1\textwidth]{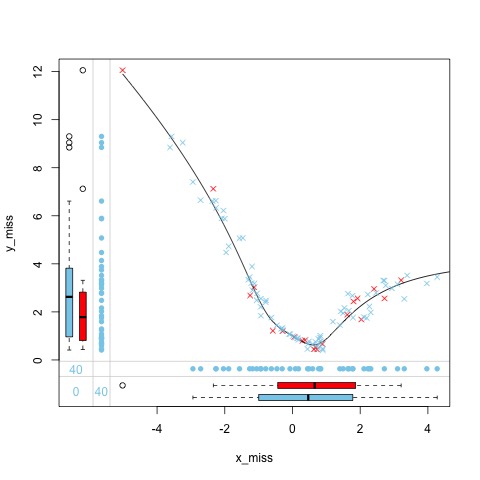}
\caption{\label{fig:mar} A generated missing data set in the simulated data case when $n=100$ and $r=0.4$. Blue crosses represent for the data items with one component missing (either $x$ or $y$) and red ones represent for the complete data items. The blue circles near the $x$-axis and the $y$-axis represent for the observed values of $x$ and $y$ respectively for the missing data items. The black curve is $E(Y|X=x)$. }
\end{figure}

We consider the two dimensional random vector $(X,Y)$. We let $X \sim N(0,2^2)$, and $$Y|X \sim N(e^{X/6}-X+\log(X^4+1),(X^2e^{-|X|})^2)$$ and we generate $(X_i,Y_i),i=1,...,n$ as our raw data set. $Y$ can be seen as a dependent variable and X as an independent variable. Then, we set a proportion $r$ of $X_i$'s to be missing completely at random (MCAR) and set $r$ of $Y_i$'s to be missing completely at random but only for those $Y_i$'s whose corresponding $X_i$ is not missing. Finally, we get the practical missing data set with sample size $n$. Table \ref{tab:snapchat} displays a missing data set we generate in this case when $n=10$ and $r=0.2$ and Figure \ref{fig:mar} displays a missing data set when $n=100$ and $r=0.4$.

\begin{table}[!htbp]
    \centering
    \begin{tabular}{c|ccccc ccccc}
  \hline
 $X$ & $x_1$ &$x_2$ & $\text{NA}$ &$x_4$ &$x_5$ &$\text{NA}$ &$x_7$ &$x_8$ &$x_9$ & $x_{10}$\\
 \hline 
 $Y$ & $\text{NA}$& $y_2$ & $y_3$& $y_4$  & $y_5$& $y_6$& $y_7$  & $y_8$& $\text{NA}$ & $y_{10}$\\ 
  \hline	
\end{tabular}
    \caption{A generated missing data set in the simulated data case when $n=10$ and $r=0.2$.}
    \label{tab:snapchat}
\end{table}

\subsubsection{Real Data Set-up}
We use the real data set ``airquality'' in R package ``mice'' and the real data set ``Iris''. In ``airquality'', we consider the variables ``Ozone'' ($X_1$), which is a dependent variable and ``Solar'' ($X_2$), ``Wind'' ($X_3$) and ``Temp'' ($X_4$), which are independent variables. Since the scale of the ``airquality'' data is large, we take $log$ for the four variables before analyzing. In ``Iris'', not considering the data of classes, there are four variables including length and width of the sepal and the petal. For both data sets, we remove the missing data items away to get the raw data set. Then, we set a proportion $r$ of each variables to be missing completely at random. If all values are missing in a data item, we remove it and then we get the practical missing data set.\\

\subsection{Measures of Performance}
To measure the performance of missing data imputation, we use the MSE criterion which measures the difference between the imputed values and true values:
$$
\text{MSE}=\sum_{i\in \Omega} \frac{(x_i^{\text{imputed}}-x_i^{\text{true}})^2}{|\Omega|}
$$
where $\Omega$ is the index set of missing data and $\lvert \cdot \rvert$ is the cardinality of the set.

To measure the performance of density estimation, we first calculate the mean estimated marginal density of each variable with respect to sampled $(\boldsymbol{\theta},\boldsymbol{\lambda})$'s and then use the Kolmogorov-Smirnov (KS) \cite{KS} method to test the goodness of fit.

\subsection{Comparison to TBMDE and Other Data Imputation Methods}
\begin{figure}
\centering
\includegraphics[width=1\textwidth]{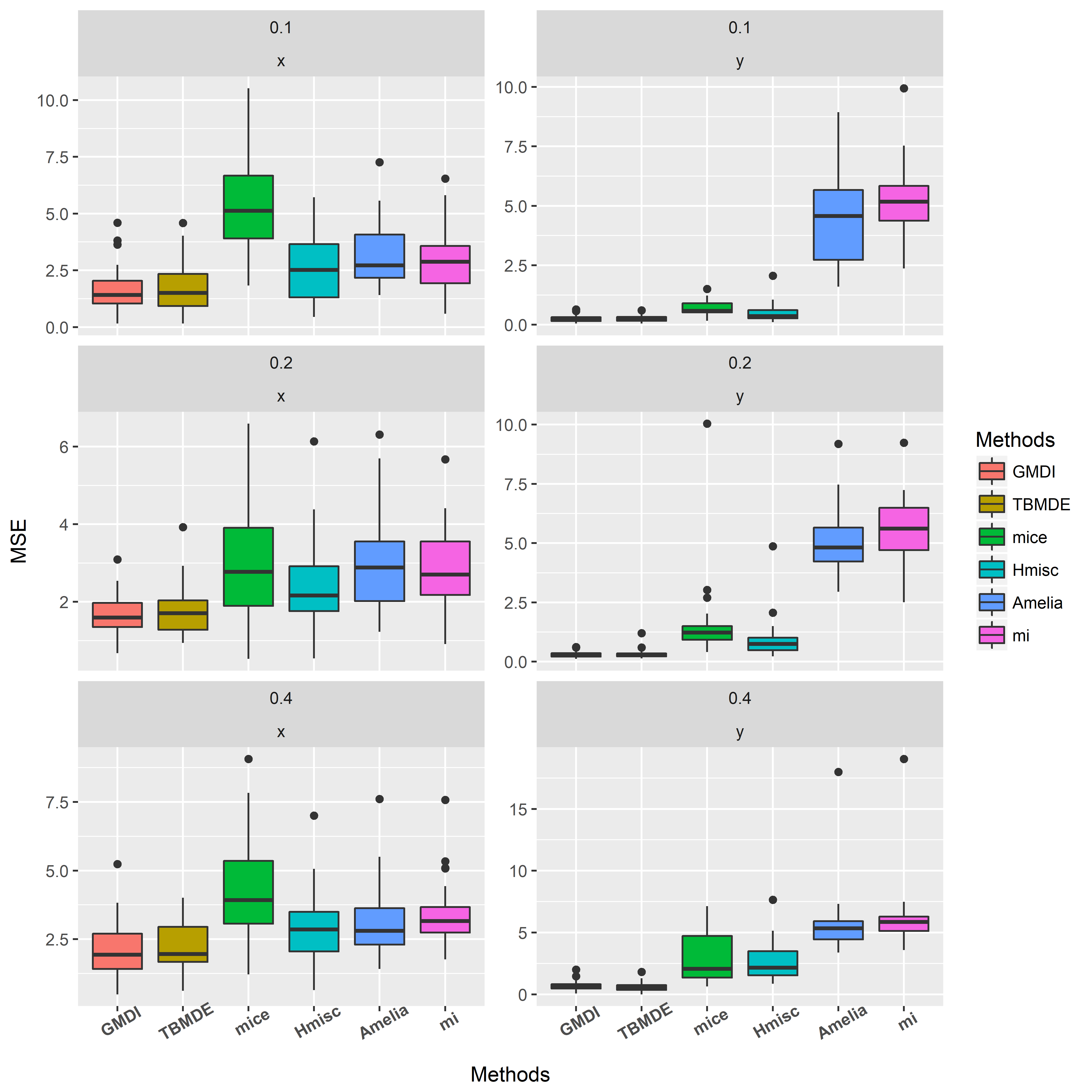}
\caption{\label{fig:2dboxplot} Boxplots of MSE for each variable in the 2 dimensional simulation data when $n=100$, with different missing proportion ($r=0.1,0.2,0.4$) and six different methods.}
\end{figure}

\begin{figure}
\centering
\includegraphics[width=1\textwidth]{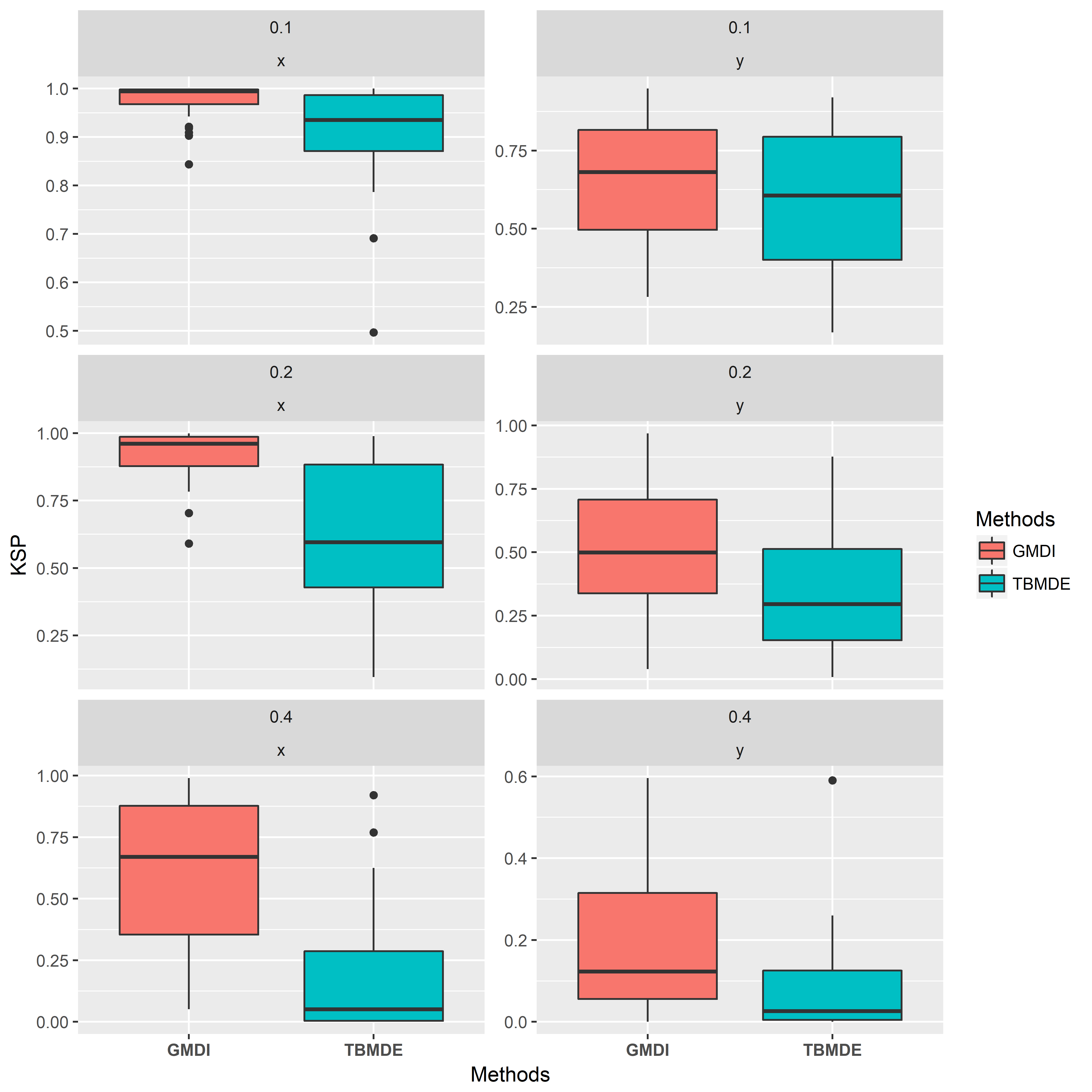}
\caption{\label{fig:ksp2dboxplot} Boxplots of KSP for each variable in the 2 dimensional simulation data when $n=100$, with different missing proportion ($r=0.1,0.2,0.4$) and two different methods: GMDI and TBMDE.}
\end{figure}

\begin{figure}
\centering
\includegraphics[width=1\textwidth]{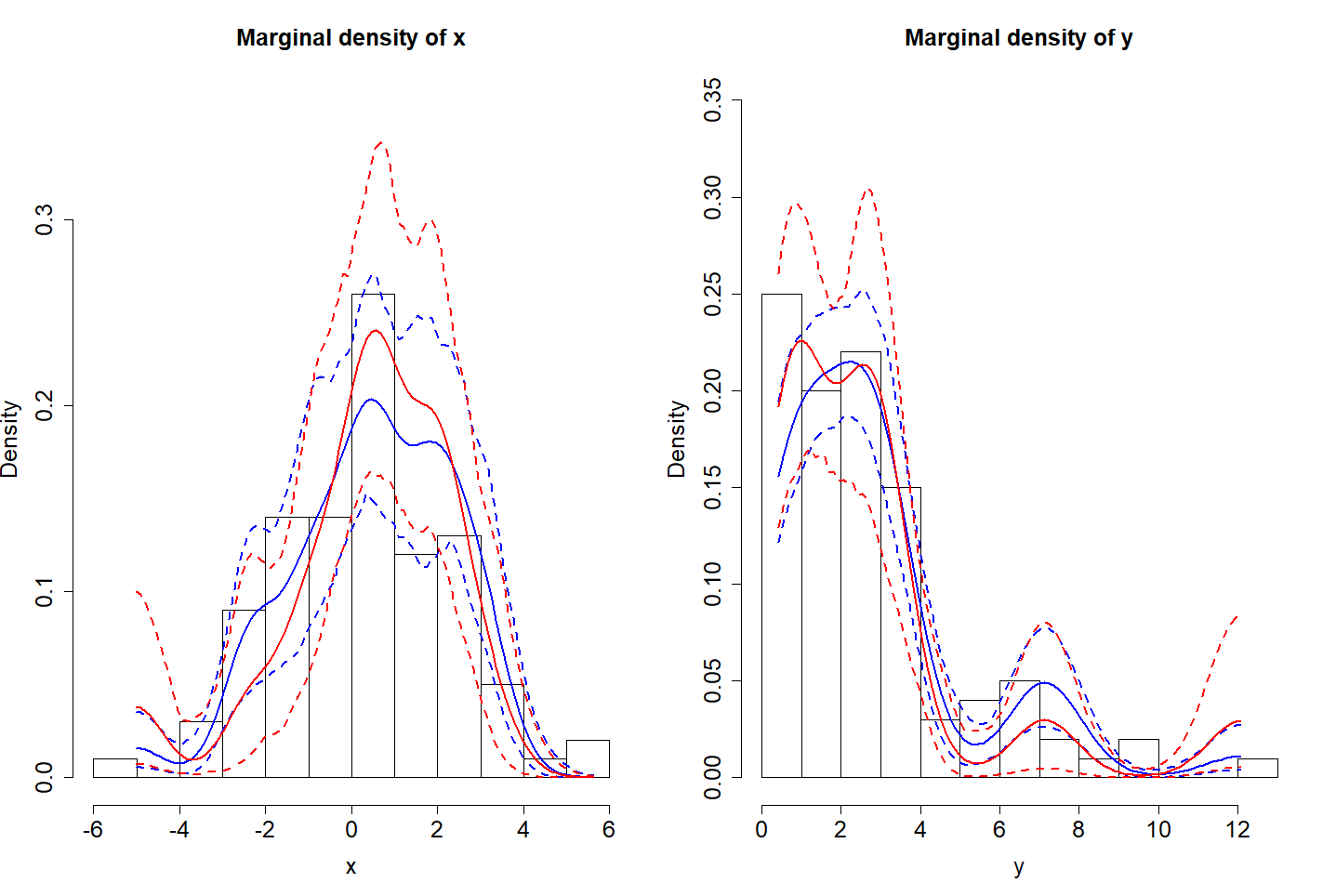}
\caption{\label{fig:MD1}Marginal density estimation of $X$ and $Y$ for the simulation data when $n=100$, $r=0.4$. Solid lines correspond to the mean estimated density while dashed lines correspond to the 2.5\% and 97.5\% quantiles of the estimated density with respect to sampled $\boldsymbol{(\theta, \lambda)}$'s. Blue and red lines correspond to density estimation using GMDI and TBMDE respectively.}
\end{figure}

Since we don't have prior information for the weights of knots, we set $\boldsymbol{\theta} \sim Dir(1/m,...,1/m)$ and $\lambda_i^2 \sim InverseGamma(n_i^{0.4}+1,b_i)$, where $n_i$ is the size of the observed data of the $i$th variable and $b_i$ is the sample variance of the $i$th variable, $i=1,...,p$. The motivation for the prior of $\lambda_i^2$ is that a rule-of-thumb estimator for the bandwidth $\lambda$ is approximately the standard deviation of the sample divided by the sample size to the one-fifth power and the mean of $InverseGamma(n_i^{0.4}+1,b_i)$ is $b_i/n_i^{0.4}$ \cite{Silverman}. For the simulation data, we set the proportion of missing values for each variable $r \in \{0.1,0.2,0.4\}$ and the number of items in a data set $n=100$. For each $(r,n)$, we impute all the missing values in the data set using GMDI, TBMDE and four other imputation methods. (The random forest imputation method from R package ``missForest" cannot be applied to two dimensional data.) As for TBMDE, we can only use the subset of the raw data where no missing value exists and we use (3) to estimate the missing values. Then we calculate the mean square error (MSE) between the missing values and the mean prediction values for each variable using the six methods. Especially for GMDI and TBMDE, we calculate the mean estimated marginal densities of $X$ and $Y$ with respect to sampled $(\boldsymbol{\theta},\boldsymbol{\lambda})$'s and use the Kolmogorov–Smirnov (KS) \cite{KS} method to test the goodness of fit. Figure \ref{fig:2dboxplot} and Figure \ref{fig:ksp2dboxplot} display the MSE and p-values in the KS test,respectively, averaged over 30 simulation data sets. Figure \ref{fig:MD1} displays the estimated marginal densities for GMDI and TBMDE for a randomly generated data set. Figure \ref{fig:PRED1} displays the true values and the predicted values in the six methods.

\begin{figure}
\centering
\includegraphics[width=1\textwidth]{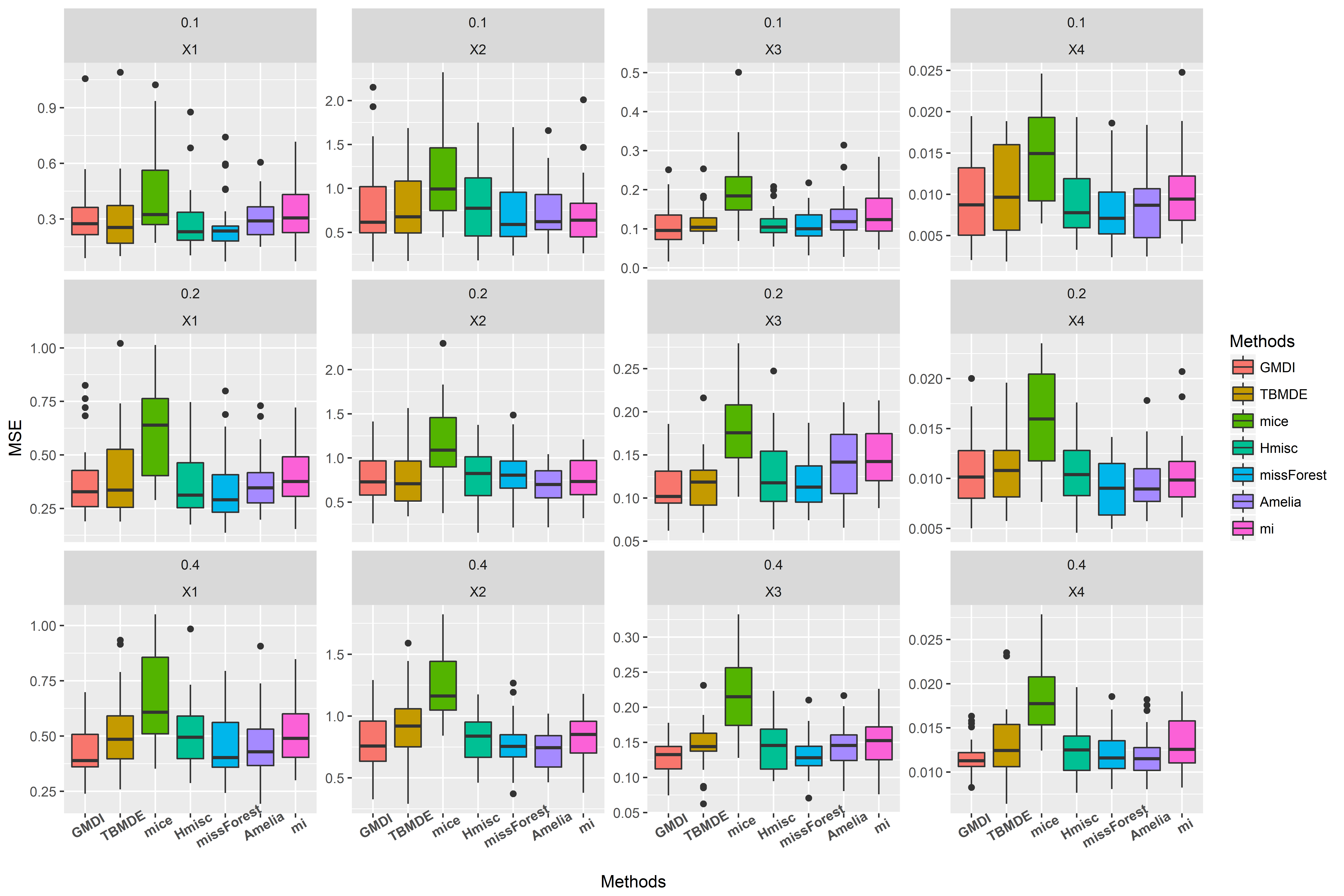}
\caption{\label{fig:airboxplot}Boxplots of MSE for each variable (take $log$ for each variable first) in the 4 dimensional ``airquality'' data with different missing proportion ($r=0.1,0.2,0.4$) and seven different methods.}
\end{figure}

\begin{figure}
\centering
\includegraphics[width=1\textwidth]{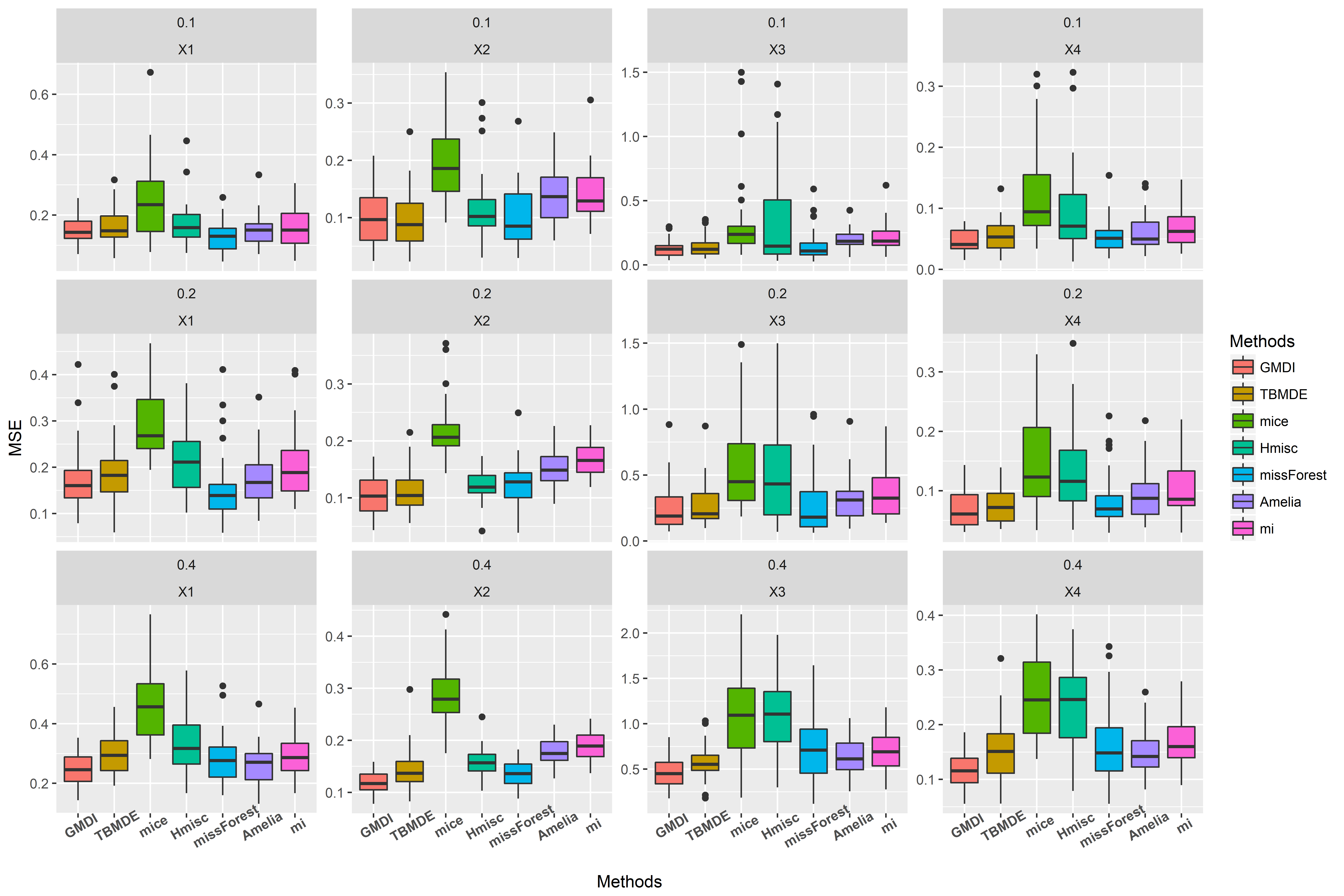}
\caption{\label{fig:irisboxplot}Boxplots of MSE for each variable in the 4 dimensional ``Iris'' data with different missing proportion ($r=0.1,0.2,0.4$) and seven different methods.}
\end{figure}

\begin{figure}
\centering
\includegraphics[width=1\textwidth]{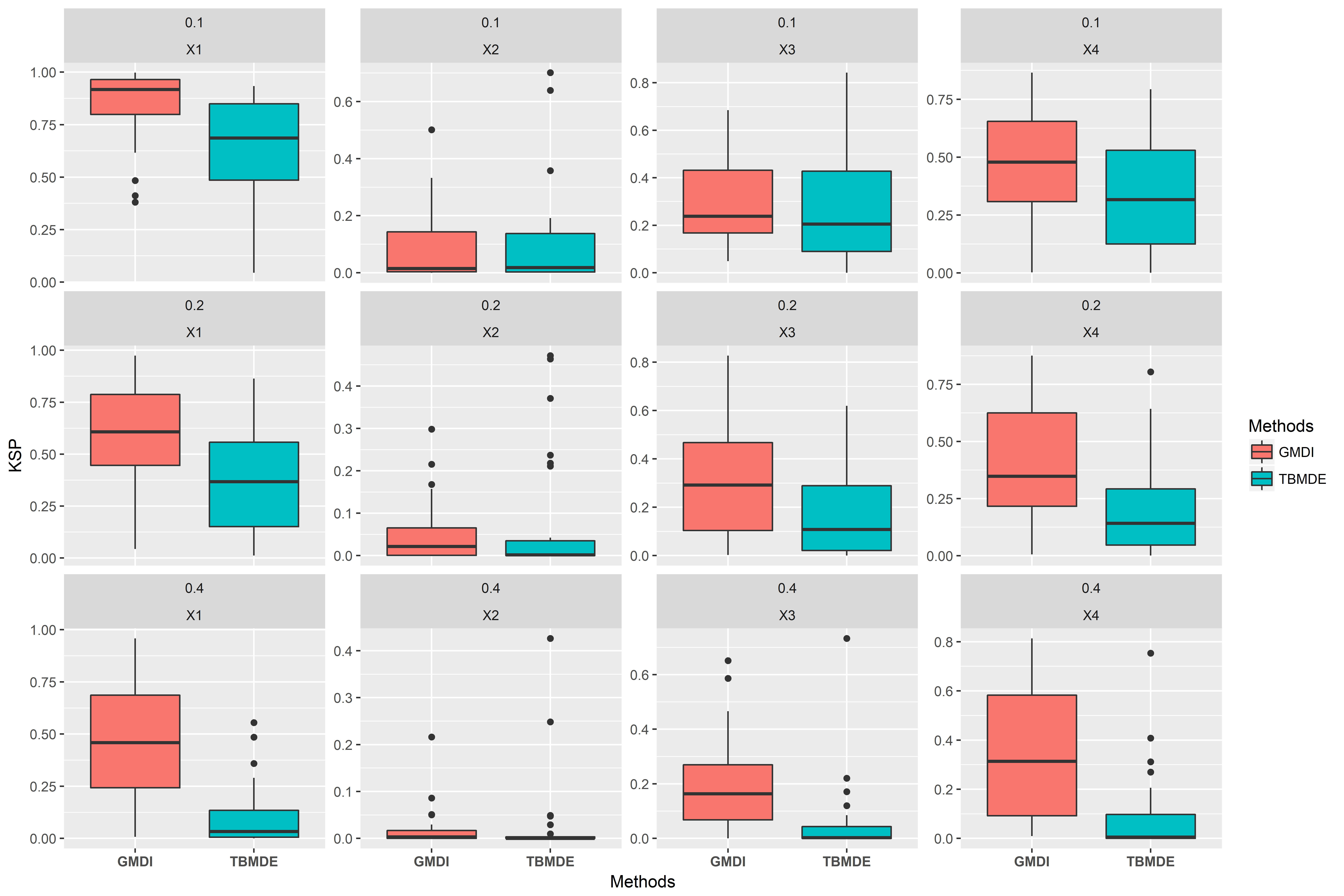}
\caption{\label{fig:kspairboxplot}Boxplots of KSP for each variable (take $log$ for each variable first) in the 4 dimensional ``airquality'' data with different missing proportion ($r=0.1,0.2,0.4$) and two different methods: GMDI and TBMDE.}
\end{figure}

\begin{figure}
\centering
\includegraphics[width=1\textwidth]{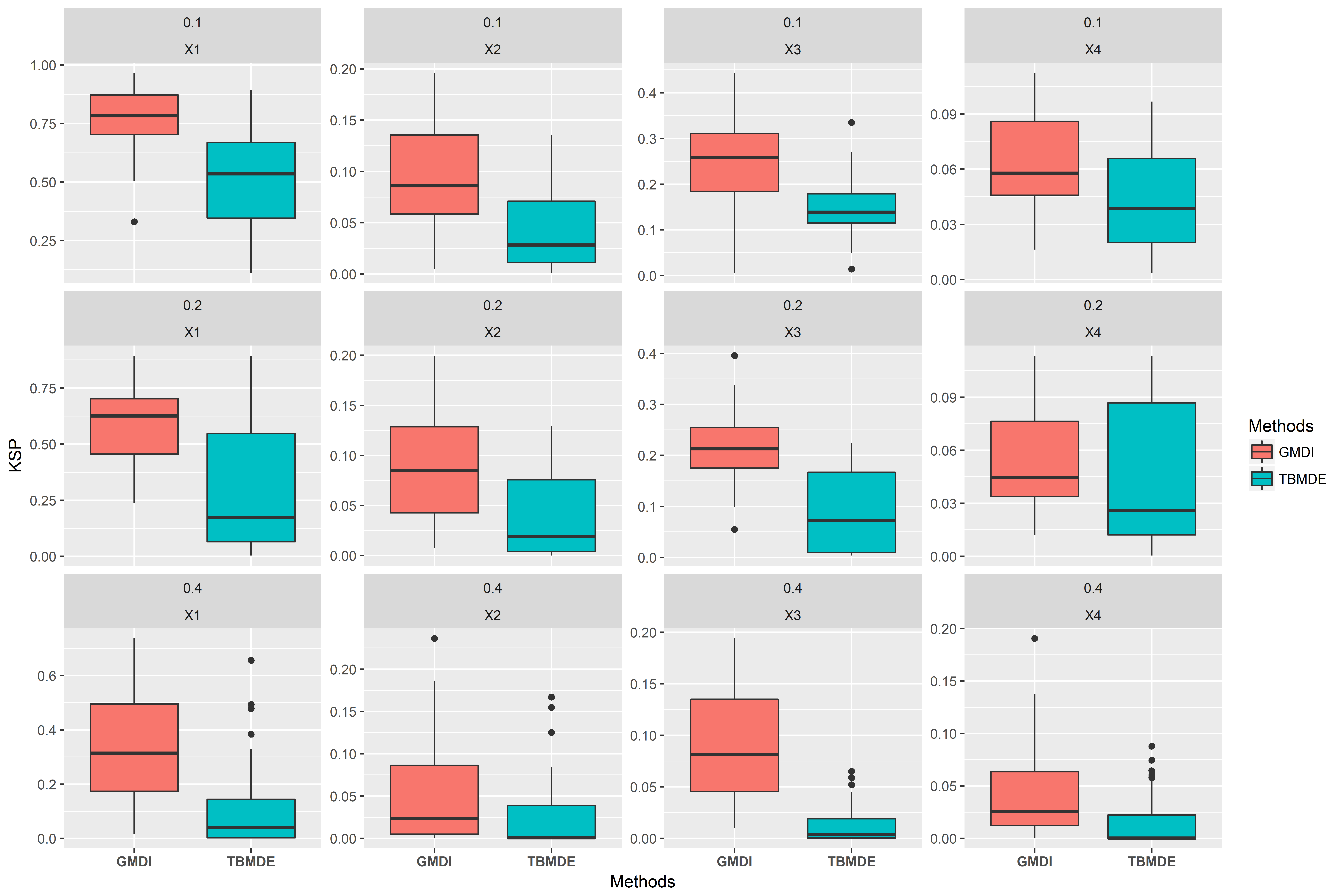}
\caption{\label{fig:kspirisboxplot}Boxplots of KSP for each variable in the 4 dimensional ``Iris'' data with different missing proportion ($r=0.1,0.2,0.4$) and two different methods: GMDI and TBMDE .}
\end{figure}

For the real data, the sample size of the raw data set is 111 for ``airquality'' and 150 for ``Iris'' and we set $r \in \{0.1,0.2,0.4 \}$. Similar as before, we impute missing values, calculate the MSE and estimate the marginal densities of $X_1,X_2,X_3,X_4$. The results of MSE are displayed in Figure \ref{fig:airboxplot}, Figure \ref{fig:irisboxplot} and the results of p-values in the KS test are displayed in Figure \ref{fig:kspairboxplot} and \ref{fig:kspirisboxplot}.

From Figure \ref{fig:2dboxplot}, \ref{fig:airboxplot} and \ref{fig:irisboxplot}, we see that the average MSE of GMDI is the smallest in many cases which shows its good performance in missing data imputation. Especially, in the four-dimensional cases, with the proportion of missing data $r$ increasing, the performance in missing data imputation of GMDI becomes better.

Figure \ref{fig:PRED1}, \ref{fig:PRED2} and \ref{fig:PRED3} show that the mean prediction values in GMDI and TBMDE are similar but GMDI has narrower empirical 95\% credible intervals. Besides, the prediction values in GMDI and TBMDE are close to the true values, especially for the response variables ($y$ in Figure \ref{fig:MD1} and $x_1$ in Figure \ref{fig:PRED2}) and rather competitive among all the methods of missing data imputation.

\begin{figure}
\centering
\includegraphics[width=1\textwidth]{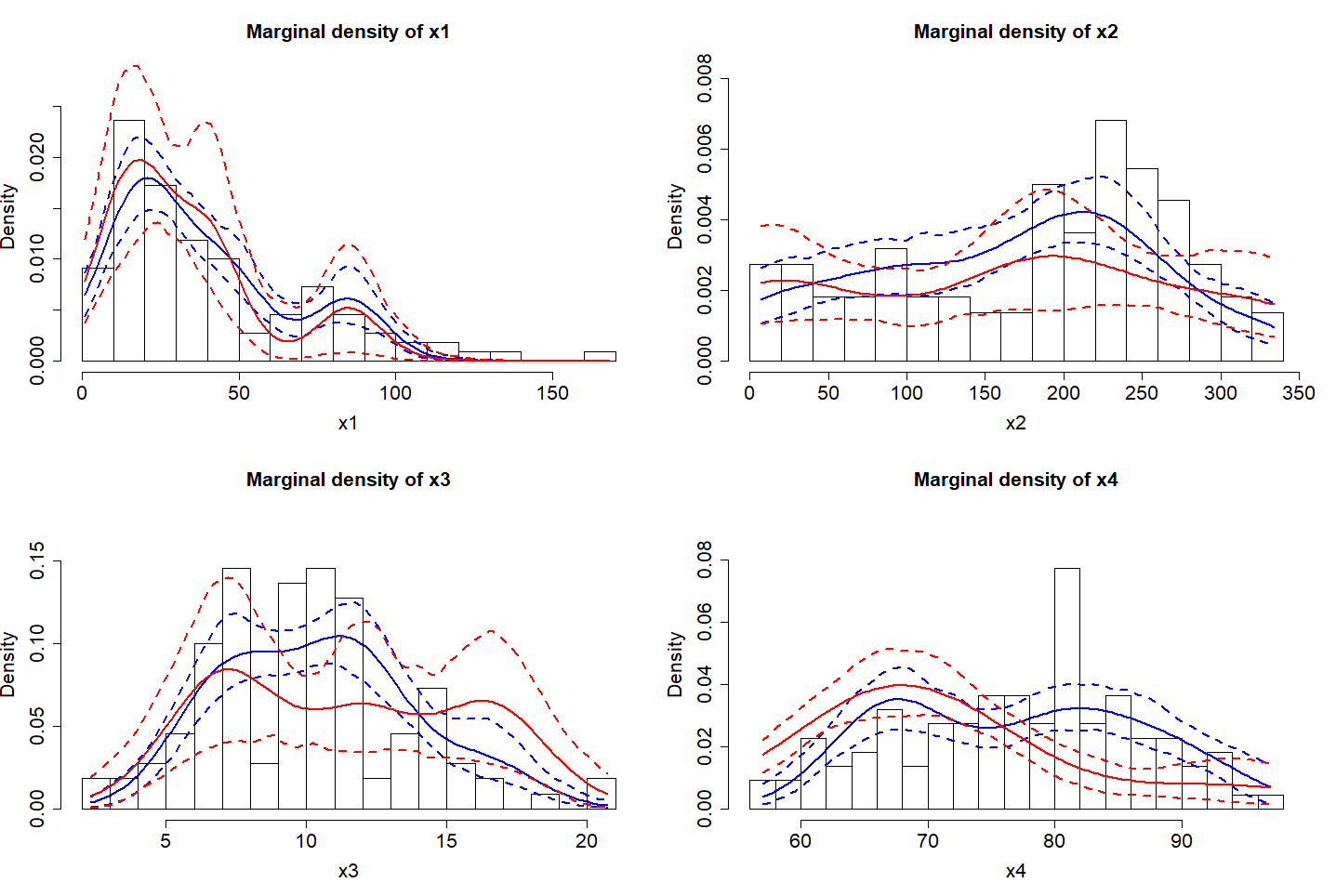}
\caption{\label{fig:MD2}Marginal density estimation of $X_1$, $X_2$, $X_3$ and $X_4$ for the ``airquality'' data when $n=111$, $r=0.4$. Solid lines correspond to the mean estimated density while dashed lines correspond to the 2.5\% and 97.5\% quantiles of the estimated density with respect to sampled $\boldsymbol{(\theta, \lambda)}$'s. Blue and red lines correspond to density estimation using GMDI and TBMDE respectively.}
\end{figure}

\begin{figure}
\centering
\includegraphics[width=1\textwidth]{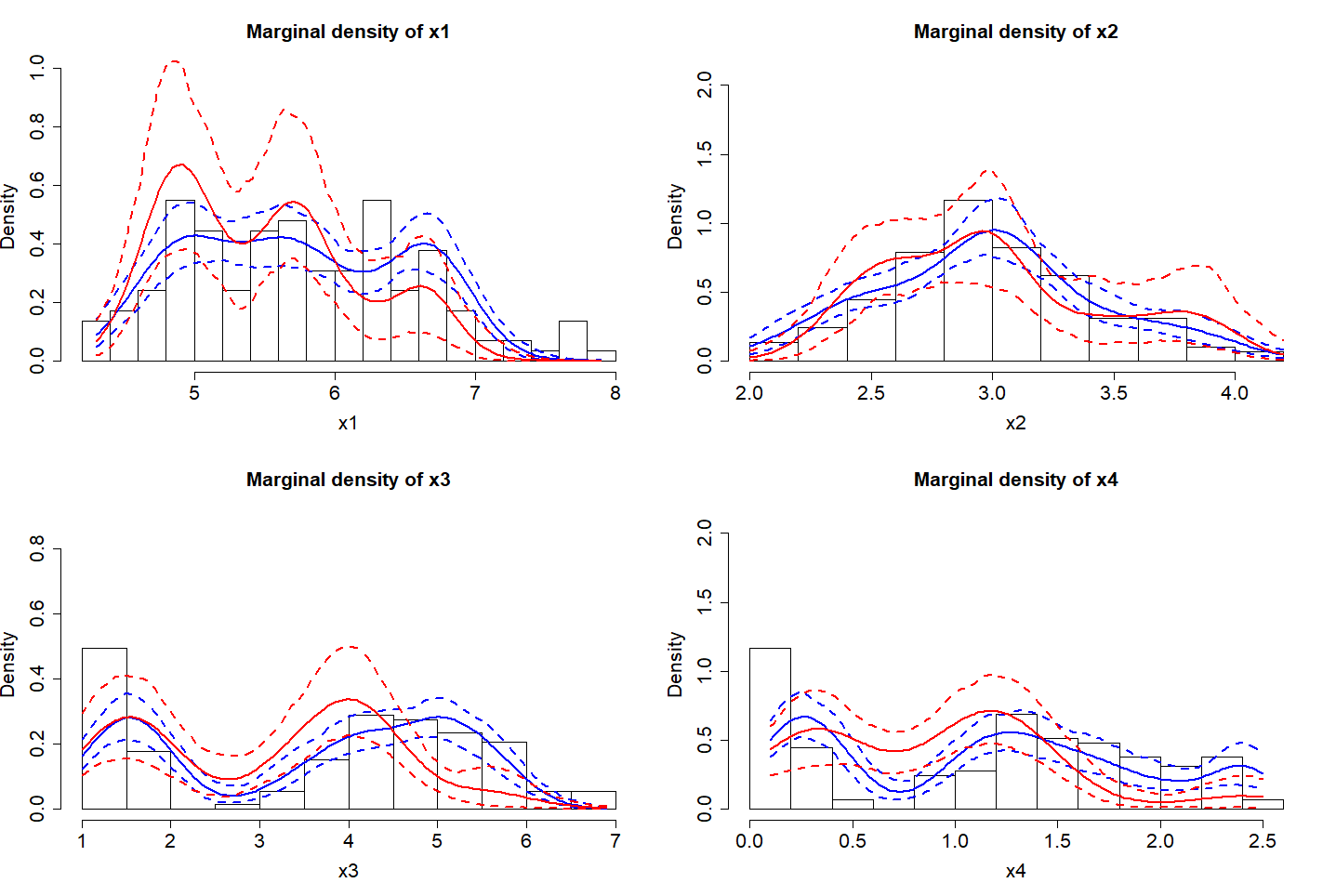}
\caption{\label{fig:MD3}Marginal density estimation of $X_1$, $X_2$, $X_3$ and $X_4$ for the ``Iris'' data when $n=150$, $r=0.4$. Solid lines correspond to the mean estimated density while dashed lines correspond to the 2.5\% and 97.5\% quantiles of the estimated density with respect to sampled $\boldsymbol{(\theta, \lambda)}$'s. Blue and red lines correspond to density estimation using GMDI and TBMDE respectively.}
\end{figure}

From Figure \ref{fig:ksp2dboxplot}, \ref{fig:kspairboxplot} and \ref{fig:irisboxplot}, we see that using the KS test, the average p-values in GMDI are larger than those in TBMDE in all cases which implies that the mean estimated joint density function in GMDI fits the data better than TBMDE. Moreover, when $n$ is fixed, with the proportion of missing data $r$ increasing, the average p-values in GMDI gradually dominate those in TBMDE. It shows that given a fixed $n$, the more missing data, the better GMDI is than TBMDE in density estimation. Figure \ref{fig:MD1}, \ref{fig:MD2} and \ref{fig:MD3} show that the mean marginal densities estimated in GMDI fit the data better than TBMDE. Moreover, the empirical 95\% credible intervals of the marginal densities in GMDI are narrower than those in TBMDE which implies the low variance of the density estimation in GMDI. 

Furthermore, Table \ref{tab:time_air} displays the average computing time of GMDI and TBMDE for the real data set $airquality$. Since GMDI makes use of both the complete data and the observed variables of the missing data, the computing time of it is longer than that of TBMDE as expected. We can also see that the computing time of GMDI does not vary much as the proportion of missing values $r$ varies while the computing time of TBMDE does.

\begin{table}[]
    \centering
    \begin{tabular}{c|c|c}
  \hline
  \hline
 r & Methods & Time(s) \\ 
  \hline
  \hline
 	\multirow{2}{*}{0.1}
    &GMDI & 	1653.20(121.09) \\ 
   & TBMDE & 820.84(93.80) \\ 
   \hline 
   \multirow{2}{*}{0.2}
   &GMDI & 1678.26(181.35)\\ 
  &TBMDE & 394.11(70.13)\\ 
   \hline
    \multirow{2}{*}{0.4}
    &GMDI& 1570.83(178.60) \\ 
  &TBMDE & 80.12(16.04) \\ 
   \hline
\end{tabular}
    \caption{Time of GMDI and TBMDE applied to the real data set (airquality).}
    \label{tab:time_air}
\end{table}

\begin{table}[]
    \centering
    \begin{tabular}{c|c|cccc}
  \hline
  \hline
 r & Methods & $x_1$ & $x_2$ & $x_3$ & $x_4$ \\ 
  \hline
  \hline
 	\multirow{7}{*}{0.1}&
GMDI & 	0.15(0.04) & 	\textbf{0.1}(0.05) & 	 \textbf{0.13}(0.07) & \textbf{	0.05}(0.02) \\ 
   & TBMDE & 0.16(0.06) & \textbf{0.1}(0.05) & 0.15(0.1) & \textbf{0.05}(0.03) \\ 
  & mice & 0.25(0.13) & 0.19(0.07) & 0.35(0.36) & 0.12(0.07) \\ 
  & Hmisc & 0.17(0.08) & 0.12(0.06) & 0.35(0.38) & 0.1(0.08) \\ 
  & missForest & 	\textbf{0.13}(0.05) & \textbf{0.1}(0.06) & 0.16(0.13) & 0.06(0.03) \\ 
  & Amelia & 0.15(0.05) & 0.14(0.05) & 0.19(0.08) & 0.06(0.03) \\ 
  & mi & 0.16(0.07) & 0.14(0.05) & 0.22(0.12) & 0.07(0.03) \\ 
   \hline 
   \multirow{7}{*}{0.2}
   &GMDI & 0.18(0.07) & 	\textbf{0.11}(0.03) & 	\textbf{0.25}(0.18) & 	\textbf{0.07}(0.03) \\ 
  &TBMDE & 0.19(0.07) & \textbf{0.11}(0.04) & 0.28(0.17) & 0.08(0.03) \\ 
  &mice & 0.3(0.08) & 0.22(0.05) & 0.57(0.35) & 0.15(0.08) \\ 
  &Hmisc & 0.21(0.07) & 0.12(0.03) & 0.51(0.37) & 0.14(0.08) \\ 
  &missForest & 	\textbf{0.16}(0.08) & 0.13(0.04) & 0.31(0.27) & 0.09(0.05) \\ 
  &Amelia & 0.18(0.06) & 0.15(0.03) & 0.32(0.18) & 0.09(0.04) \\ 
  &mi & 0.21(0.08) & 0.17(0.03) & 0.38(0.21) & 0.1(0.05) \\ 
   \hline
    \multirow{7}{*}{0.4}
    &GMDI& \textbf{0.25}(0.06) & 	\textbf{0.12}(0.02) & 	\textbf{0.47}(0.18) & 	\textbf{0.12}(0.03) \\ 
  &TBMDE & 0.3(0.07) & 0.15(0.04) & 0.58(0.2) & 0.16(0.06) \\ 
  &mice & 0.46(0.12) & 0.29(0.06) & 1.13(0.47) & 0.25(0.09) \\ 
  &Hmisc & 0.33(0.1) & 0.16(0.03) & 1.08(0.42) & 0.23(0.08) \\ 
  &missForest & 0.29(0.09) & 0.14(0.03) & 0.74(0.37) & 0.16(0.07) \\ 
  &Amelia & 0.26(0.07) & 0.18(0.03) & 0.64(0.21) & 0.15(0.04) \\ 
  &mi & 0.29(0.07) & 0.19(0.03) & 0.72(0.23) & 0.17(0.05) \\ 

   \hline
\end{tabular}
    \caption{MSE of true data and prediction in GMDI, TBMDE and the other five R packages for the real dataset (Iris). The bold values are the best ones in each column.}
    \label{tab:iris}
\end{table}

\begin{table}[]
    \centering
    \begin{tabular}{c|c|cc}
    \hline
    \hline
   r & Variable & KSP(GMDI) & KSP(TBMDE)  \\
    \hline
    \hline
    \multirow{4}{*}{0.1}
   & $x_1$ & 	\textbf{0.76}(0.15) & 0.51(0.22) \\
  &$x_2$ & 	\textbf{0.09}(0.06) & 0.04(0.04) \\ 
  &$x_3$ & 	\textbf{0.25}(0.11) & 0.15(0.07) \\ 
  &$x_4$ & 	\textbf{0.06}(0.03) & 0.04(0.03) \\ 
    \hline
    \multirow{4}{*}{0.2}
      & $x_1$ &  	\textbf{0.59}(0.17) & 0.3(0.27) \\ 
   &$x_2$  & 	\textbf{0.09}(0.06) & 0.04(0.04) \\ 
  &$x_3$& 	\textbf{0.22}(0.08) & 0.09(0.08) \\ 
  &$x_4$ & \textbf{0.06}(0.03) & 0.04(0.04) \\ 
    \hline
    \multirow{4}{*}{0.4}
        & $x_1$ &  	\textbf{0.36}(0.21) & 0.12(0.18) \\ 
   &$x_2$  & 	\textbf{0.05}(0.06) & 0.03(0.05) \\ 
  &$x_3$& 	\textbf{0.09}(0.06) & 0.01(0.02) \\ 
  &$x_4$ & \textbf{0.04}(0.05) & 0.02(0.03) \\ 
    \hline

    \end{tabular}
    \caption{KS test p-values of density estimation by GMDI and TBMDE for the real dataset (Iris). The bold values are the best ones in each row.}
    \label{tab:iris_ks}
\end{table}

\newpage
\section{Discussion}
In our method, we make use of data with missing values to do density estimation and impute missing values. From the aspect of imputing missing data, GMDI performs better in prediction than some traditional imputation method like PMM. From the aspect of density estimation, GMDI performs better than some traditional density estimation method like TBMDE which cannot use missing data. However, since GMDI uses three Gibbs samplers, its computational complexity is larger than TBMDE so it's not time efficient. Moreover, the cross-validation method we use is based on MSE criterion and it costs much time to implement. In the future, we can consider the BIC, DIC or Bayesian cross-validation for tuning parameters selection. Besides, we can also give the number of knots $m$ a certain prior in order to implement a full Bayesian method.

\section*{References}

\bibliography{mybibfile}

\appendix
\section{\\Appendix}
If we do not consider prior on $\boldsymbol{\lambda}$, we can use the following empirical procudure to choose bandwidth $\lambda_i, i=1,...,p$:  $\lambda_i$ is the bandwidth of variable $X_i$. We implement a rule-of-thumb for choosing the bandwidth of a Gaussian kernel density estimator. It defaults to 0.9 times the minimum of the standard deviation and the interquartile range divided by 1.34 times the sample size to the negative one-fifth power (i.e. Silverman's "rule of thumb", Silverman (1986))\cite{Silverman};

\section{Gibbs Sampler Review}
Gibbs sampling \cite{Casella} is a special case of the Metropolis–Hastings algorithm. The point of Gibbs sampling is that given a multivariate distribution it is simpler to sample from a conditional distribution than to marginalize by integrating over a joint distribution. Suppose we want to obtain $k$ samples of $\boldsymbol{\theta}=(\theta_1,\theta_2,...,\theta_m)$ from a joint distribution $P(\theta_1,\theta_2,...,\theta_{m})$. We denote the $i$th sample of them as $\boldsymbol{\theta}^{(i)}=(\theta^{(i)}_1,\theta^{(i)}_2,...,\theta^{(i)}_m)$. We proceed as follows:
we begin with some initial value $\boldsymbol{\theta}^{(0)}$ and suppose $p(\theta_j |\theta_{-j} )$ denotes the conditional density of $\theta_j$ given $\{\theta_k: k\notin j,1\leq k \leq m\}$. If we get the sample $\theta^{(k)}$,  we get the sample $\theta^{(k+1)}$ for each component of it under the mechanism $$\theta^{(k+1)}_j\sim p(\theta^{(k+1)}_j|\theta^{(k+1)}_1,...,\theta^{(k+1)}_{j-1},\theta^{(k)}_{j+1},...,\theta^{(k)}_m).$$\\
We repeat this step for $m$ times and we will get a renewed version of $\theta^{(k+1)}$\\
If such sampling is performed, these important facts hold:
\begin{itemize}
\item The samples approximate the joint distribution of all variables.
\item The marginal distribution of any subset of variables can be approximated by simply considering the samples for that subset of variables, ignoring the rest.
\item The expected value of any variable can be approximated by averaging over all the samples.
\end{itemize}

\begin{algorithm}[H]
        \begin{algorithmic}[1]
            \STATE Suppose we have $p$ dimensional complete data $\mathbf{x}_1,...\mathbf{x}_n$, $f(\mathbf{x}|\boldsymbol{\theta},\boldsymbol{\lambda})=\sum_{k=1}^m \theta_k f_k(\mathbf{x}|\boldsymbol{\lambda})$.\\ $\boldsymbol{\theta} \sim Dir(\alpha_1,..., \alpha_m)$ and $\boldsymbol{\lambda}$ is fixed.
            \STATE Initialize $\boldsymbol{\theta}^{(0)}=(\frac{1}{m},\frac{1}{m},...,\frac{1}{m})$. 
            \FOR {iteration $l=1,2,...$} 
            \STATE Sample $K_i^{(l)}\sim \frac{\theta_k^{(l-1)}f_k(\mathbf{x}_i|\boldsymbol{\lambda})}{\sum_{k=1}^{m}\theta_k^{(l-1)}f_k(\mathbf{x}_i|\boldsymbol{\lambda})}I(k_i=k)$, for $i=1,2,...,n$.
            \STATE Sample $\boldsymbol{\theta}^{(l)} \sim Dir(n_1(K^{(l)})+a_1,...,n_m(K^{(l)})+a_m)$, where $n_k(K^{(l)})=\sum_{i=1}^nI(K_i^{(l)}=k)$, for $k=1,...,m$.
            \ENDFOR
            \end{algorithmic}
            \caption{Sample $\boldsymbol{\theta}$ given $\boldsymbol{\lambda}$}
            \label{alg:condTheta}
            \end{algorithm}
            
Derivation of Algorithm 3:
$$
\begin{aligned}
p(\boldsymbol{\theta}'|\mathbf{x_1},...,\mathbf{x_n},\boldsymbol{\lambda})&=cp(\mathbf{x_1},...,\mathbf{x_n}|\boldsymbol{\theta}',\boldsymbol{\lambda})p(\boldsymbol{\theta}'|\boldsymbol{\lambda})\\
&=c\prod_{i=1}^n f(\mathbf{x_i}|\boldsymbol{\theta}',\boldsymbol{\lambda})p(\boldsymbol{\theta}'|\boldsymbol{\lambda})\\
&=c\prod_{i=1}^n \sum_{k=1}^m \theta_k' f_k(\mathbf{x_i}|\boldsymbol{\lambda}) p(\boldsymbol{\theta}')\\
&=c'\prod_{i=1}^n \sum_{k=1}^m \theta_k' f_k(\mathbf{x_i}|\boldsymbol{\lambda}) \prod_{i=1}^m \theta_i'^{a_i-1},
\end{aligned}
$$
where $c$ and $c'$ are constants not related to $\boldsymbol{\theta}'$.

The transition kernel density (TKD) of the Markov chain with respect to $\boldsymbol{\theta}$ is
$$
\begin{aligned}
T(\boldsymbol{\theta}',\boldsymbol{\theta})&=\sum_{\mathbf{k}} p(\boldsymbol{\theta}|\mathbf{k},\mathbf{x_1},...,\mathbf{x_n},\boldsymbol{\lambda}) p(\mathbf{k}|\boldsymbol{\theta}',\mathbf{x_1},...,\mathbf{x_n},\boldsymbol{\lambda})\\
&=\sum_{\mathbf{k}} (\frac{1}{B(\mathbf{a})}\prod_{i=1}^m \theta_i^{n_i(\mathbf{k})+a_i-1} \prod_{i=1}^n \frac{\theta_{k_i}'f_{k_i}(\mathbf{x_i}|\boldsymbol{\lambda})}{\sum_{k=1}^m \theta_k'f_k(\mathbf{x_i}|\boldsymbol{\lambda})})
\end{aligned}
$$
where $\mathbf{k}=(k_1,...,k_n)'$, $k_i \in \{1,...,m\},i=1,...,n$ and $\mathbf{a}=(n_1(\mathbf{k})+a_1,...,n_m(\mathbf{k})+a_m)'$, $B(\mathbf{y})=\frac{\prod_{i=1}^m \Gamma(y_i)}{\Gamma(\sum_{i=1}^m y_i)}$.

Then,
$$
\begin{aligned}
p(\boldsymbol{\theta}'|\mathbf{x_1},...,\mathbf{x_n},\boldsymbol{\lambda})T(\boldsymbol{\theta}',\boldsymbol{\theta})&=c'\prod_{i=1}^n \sum_{k=1}^m \theta_k' f_k(\mathbf{x_i}|\boldsymbol{\lambda}) \prod_{i=1}^m \theta_i'^{a_i-1} \sum_{\mathbf{k}} (\frac{1}{B(\mathbf{a})}\prod_{i=1}^m \theta_i^{n_i(\mathbf{k})+a_i-1} \prod_{i=1}^n \frac{\theta_{k_i}'f_{k_i}(\mathbf{x_i}|\boldsymbol{\lambda})}{\sum_{k=1}^m \theta_k'f_k(\mathbf{x_i}|\boldsymbol{\lambda})})\\
&=\frac{c'}{B(\mathbf{a})}\prod_{i=1}^m \theta_i'^{a_i-1}\sum_{\mathbf{k}}(\prod_{i=1}^m \theta_i^{n_i(\mathbf{k})+a_i-1} \prod_{i=1}^n \theta'_{k_i}f_{k_i}(\mathbf{x_i}|\boldsymbol{\lambda}))\\
&=\frac{c'}{B(\mathbf{a})} \sum_{\mathbf{k}} (\prod_{i=1}^m \theta_i^{n_i(\mathbf{k})+a_i-1} \prod_{i=1}^m \theta_i'^{n_i(\mathbf{k})+a_i-1}) \prod_{i=1}^n f_{k_i}(\mathbf{x_i}|\boldsymbol{\lambda})\\
&=p(\boldsymbol{\theta}|\mathbf{x_1},...,\mathbf{x_n},\boldsymbol{\lambda})T(\boldsymbol{\theta},\boldsymbol{\theta}').
\end{aligned}
$$
The last equality holds since the expression in the second to last line is symmetric with respect to $\boldsymbol{\theta}$ and $\boldsymbol{\theta}'$.

Hence,
$$
\begin{aligned}
p(\boldsymbol{\theta}|\mathbf{x_1},...,\mathbf{x_n},\boldsymbol{\lambda})&=\int_{\theta'} p(\boldsymbol{\theta}|\mathbf{x_1},...,\mathbf{x_n},\boldsymbol{\lambda})T(\boldsymbol{\theta},\boldsymbol{\theta}') d \boldsymbol{\theta}'\\
&=\int_{\boldsymbol{\theta}'} p(\boldsymbol{\theta}'|\mathbf{x_1},...,\mathbf{x_n},\boldsymbol{\lambda})T(\boldsymbol{\theta}',\boldsymbol{\theta}) d \boldsymbol{\theta}',
\end{aligned}
$$
which shows the density $p(\boldsymbol{\theta}|\mathbf{x_1},...,\mathbf{x_n},\boldsymbol{\lambda})$ is stationary for the transition kernel density $T(\boldsymbol{\theta}',\boldsymbol{\theta})$ so the MCMC algorithm to sample $\boldsymbol{\theta}$ given $\boldsymbol{\lambda}$ works.

\begin{algorithm}[H]
        \begin{algorithmic}[1]
            \STATE Suppose we have $p$ dimensional complete data $\mathbf{x}_1,...\mathbf{x}_n$, $f(\mathbf{x}|\boldsymbol{\theta},\boldsymbol{\lambda})=\sum_{k=1}^m \theta_k f_k(\mathbf{x}|\boldsymbol{\lambda})$.\\ $\lambda_i^2 \sim InverseGamma(a_i,b_i),\;i=1,...,p$, $\{\lambda_i\}_{1\leq i \leq p}$ are independent and $\boldsymbol{\theta}$ is fixed.
            \STATE Initialize $\boldsymbol{\lambda}^{2(0)}$. 
            \FOR {iteration $l=1,2,...$} 
            \STATE Sample $K_i^{(l)}\sim \frac{\theta_k f_k(\mathbf{x}_i|\boldsymbol{\lambda}^{(l-1)})}{\sum_{k=1}^{m}\theta_k f_k(\mathbf{x}_i|\boldsymbol{\lambda}^{(l-1)})}I(k_i=k)$, for $i=1,2,...,n$.
            \STATE Sample $\lambda_i^{2(l)} \sim InverseGamma(\frac{n}{2}+a_i,\frac{\sum_{j=1}^n (x_{ji}-s_{ik_{j}})^2}{2}+b_i)$, for $i=1,...,p$, where $x_{ji}$ denotes the $i^{th}$ variable of $\mathbf{x}_j$.
            \ENDFOR
            \end{algorithmic}
            \caption{Sample $\boldsymbol{\lambda}$ given $\boldsymbol{\theta}$}
            \label{alg:condLambda}
            \end{algorithm}
            
Derivation of Algorithm 4:
$$
\begin{aligned}
p(\boldsymbol{\lambda}'^2|\mathbf{x_1},...,\mathbf{x_n},\boldsymbol{\theta})&=cp(\mathbf{x_1},...,\mathbf{x_n}|\boldsymbol{\theta},\boldsymbol{\lambda}'^2)p(\boldsymbol{\lambda}'^2|\boldsymbol{\theta})\\
&=c\prod_{i=1}^n f(\mathbf{x_i}|\boldsymbol{\theta},\boldsymbol{\lambda}'^2)p(\boldsymbol{\lambda}'^2|\boldsymbol{\theta})\\
&=c\prod_{i=1}^n \sum_{k=1}^m \theta_k f_k(\mathbf{x_i}|\boldsymbol{\lambda}'^2) p(\boldsymbol{\lambda}'^2)\\
&=c'\prod_{i=1}^n \sum_{k=1}^m \theta_k f_k(\mathbf{x_i}|\boldsymbol{\lambda}'^2) \prod_{i=1}^p (\lambda_i^{2(-a_i-1)}e^{-\frac{b_i}{\lambda_i^2}}),
\end{aligned}
$$
where $c$ and $c'$ are constants not related to $\boldsymbol{\lambda}'^2$.

The transition kernel density (TKD) of the Markov chain with respect to $\boldsymbol{\lambda}^2$ is
$$
\begin{aligned}
T(\boldsymbol{\lambda}'^2,\boldsymbol{\lambda}^2)&=\sum_{\mathbf{k}} p(\boldsymbol{\lambda}^2|\mathbf{k},\mathbf{x_1},...,\mathbf{x_n},\boldsymbol{\theta}) p(\mathbf{k}|\boldsymbol{\theta},\mathbf{x_1},...,\mathbf{x_n},\boldsymbol{\lambda}'^2)\\
&=\sum_{\mathbf{k}} (\prod_{i=1}^p \frac{b_i(\mathbf{k})^{n/2+a_i}}{\Gamma(n/2+a_i)} \lambda_i^{-2(n/2+a_i+1)} e^{-\frac{b_i(\mathbf{k})}{\lambda_i^2}} \prod_{i=1}^n \frac{\theta_{k_i}f_{k_i}(\mathbf{x_i}|\boldsymbol{\lambda}'^2)}{\sum_{k=1}^m \theta_k f_k(\mathbf{x_i}|\boldsymbol{\lambda}'^2)})
\end{aligned}
$$
where $\mathbf{k}=(k_1,...,k_n)'$, $k_i \in \{1,...,m\},i=1,...,n$ and $b_i(\mathbf{k})=\frac{\sum_{j=1}^n (x_{ji}-s_{ik_{j}})^2}{2}+b_i,i=1,...,p$.

Then,
$$
\begin{aligned}
p(\boldsymbol{\lambda}'^2|\mathbf{x_1},...,\mathbf{x_n},\boldsymbol{\theta})T(\boldsymbol{\lambda}'^2,\boldsymbol{\lambda}^2)&=c'\prod_{j=1}^n \sum_{k=1}^m \theta_k f_k(\mathbf{x_j}|\boldsymbol{\lambda}'^2) \prod_{i=1}^p (\lambda_i'^{2(-a_i-1)}e^{-\frac{b_i}{\lambda_i'^2}})\\ 
& \times \sum_{\mathbf{k}} (\prod_{i=1}^p \frac{b_i(\mathbf{k})^{n/2+a_i}}{\Gamma(n/2+a_i)} \lambda_i^{-2(n/2+a_i+1)} e^{-\frac{b_i(\mathbf{k})}{\lambda_i^2}} \prod_{j=1}^n \frac{\theta_{k_j}f_{k_j}(\mathbf{x_j}|\boldsymbol{\lambda}')}{\sum_{k=1}^m \theta_k f_k(\mathbf{x_j}|\boldsymbol{\lambda}')})\\
&=c'\prod_{i=1}^p (\lambda_i'^{2(-a_i-1)}e^{-\frac{b_i}{\lambda_i'^2}}) \sum_{\mathbf{k}} (\prod_{i=1}^p \frac{b_i(\mathbf{k})^{n/2+a_i}}{\Gamma(n/2+a_i)} \lambda_i^{-2(n/2+a_i+1)} e^{-\frac{b_i(\mathbf{k})}{\lambda_i^2}} \\
&\times \prod_{j=1}^n \theta_{k_j}f_{k_j}(\mathbf{x_j}|\boldsymbol{\lambda}'))\\
&=c'\prod_{i=1}^p (\lambda_i'^{2(-a_i-1)}e^{-\frac{b_i}{\lambda_i'^2}}) \sum_{\mathbf{k}} (\prod_{i=1}^p \frac{b_i(\mathbf{k})^{n/2+a_i}}{\Gamma(n/2+a_i)} \lambda_i^{-2(n/2+a_i+1)} e^{-\frac{b_i(\mathbf{k})}{\lambda_i^2}} \\
&\times \prod_{j=1}^n \theta_{k_j} \prod_{i=1}^p \frac{1}{\sqrt{2\pi}\lambda_i'}e^{-\frac{(x_{ji}-s_{ik_j})^2}{2\lambda_i'^2}})\\
&=c'\sum_{\mathbf{k}} (\prod_{i=1}^p (\frac{b_i(\mathbf{k})^{n/2+a_i}}{\Gamma(n/2+a_i)} \lambda_i^{-2(n/2+a_i+1)} e^{-\frac{b_i(\mathbf{k})}{\lambda_i^2}} \lambda_i'^{-2(n/2+a_i+1)} \\
&\times e^{-\sum_{j=1}^n \frac{(x_{ji}-s_{ik_j})^2}{2\lambda_i'^2}-\frac{b_i}{\lambda_i'^2}}) \prod_{j=1}^n \theta_{k_j} \prod_{i=1}^p \frac{1}{\sqrt{2\pi}} )\\
&=c'\sum_{\mathbf{k}} (\prod_{i=1}^p (\frac{b_i(\mathbf{k})^{n/2+a_i}}{\Gamma(n/2+a_i)} \lambda_i^{-2(n/2+a_i+1)} e^{-\frac{b_i(\mathbf{k})}{\lambda_i^2}} \lambda_i'^{-2(n/2+a_i+1)} \\
&\times e^{-\frac{b_i(\mathbf{k})}{\lambda_i'^2}}) \prod_{j=1}^n \theta_{k_j} \prod_{i=1}^p \frac{1}{\sqrt{2\pi}} )\\
&=p(\boldsymbol{\lambda^2}|\mathbf{x_1},...,\mathbf{x_n},\boldsymbol{\theta})T(\boldsymbol{\lambda}^2,\boldsymbol{\lambda}'^2).
\end{aligned}
$$
The last equality holds since the expression in the second to last line is symmetric with respect to $\boldsymbol{\lambda}^2$ and $\boldsymbol{\lambda}'^2$.

Hence,
$$
\begin{aligned}
p(\boldsymbol{\lambda}^2|\mathbf{x_1},...,\mathbf{x_n},\boldsymbol{\theta})&=\int_{\lambda'^2} p(\boldsymbol{\lambda}^2|\mathbf{x_1},...,\mathbf{x_n},\boldsymbol{\theta})T(\boldsymbol{\lambda}^2,\boldsymbol{\lambda}'^2) d \boldsymbol{\lambda}'^2\\
&=\int_{\boldsymbol{\lambda}'^2} p(\boldsymbol{\lambda}'^2|\mathbf{x_1},...,\mathbf{x_n},\boldsymbol{\theta})T(\boldsymbol{\lambda}'^2,\boldsymbol{\lambda}^2) d \boldsymbol{\lambda}'^2,
\end{aligned}
$$
which shows the density $p(\boldsymbol{\lambda}^2|\mathbf{x_1},...,\mathbf{x_n},\boldsymbol{\theta})$ is stationary for the transition kernel density $T(\boldsymbol{\lambda}'^2,\boldsymbol{\lambda}^2)$ so the MCMC algorithm to sample $\boldsymbol{\lambda}^2$ given $\boldsymbol{\theta}$ works.

\begin{table}[]
    \centering
    \begin{tabular}{c|c|cc}
    \hline
    \hline
    r  &  Variable & KSP(GMDI) & KSP(TBMDE)    \\
    \hline
    \hline
    \multirow{2}{*}{0.1}
    & $x$ &  \textbf{0.97}(0.04) & 0.91(0.11) \\ 
     &$y$ & \textbf{0.66}(0.2) & 0.59(0.22) \\
     
     \hline
      \multirow{2}{*}{0.2}
    & $x$ & \textbf{0.92}(0.1) & 0.63(0.26) \\  
    &  $y$ & \textbf{0.51}(0.24) & 0.34(0.23) \\
  
    \hline
    \multirow{2}{*}{0.4} 
    & $x$ & \textbf{0.61}(0.3) & 0.19(0.27) \\ 
      &$y$ & \textbf{0.19}(0.18) & 0.08(0.13) \\ 
  
    \hline
    \end{tabular}
    \caption{KS test p-values of density estimation by GMDI and TBMDE for the simulation data ($n=100$). The bold values are the best ones in each row.}
    \label{tab:sim_data_ks}
\end{table}

\begin{table}[]
    \centering
    \begin{tabular}{c|c|cc}
  \hline
  \hline
 r & Methods & $x$ & $y$\\ 
  \hline
  \hline
 	\multirow{6}{*}{0.1}&
GMDI & 	\textbf{1.68}(1.06) & 	\textbf{0.25}(0.14) \\ 
  &TBMDE & 1.71(1.09) & 0.26(0.13) \\ 
  &mice & 5.45(2.16) & 0.68(0.32) \\ 
  &Hmisc & 2.65(1.54) & 0.5(0.39) \\ 
  &Amelia & 3.19(1.44) & 4.54(1.83) \\ 
  &mi & 2.94(1.51) & 5.24(1.46) \\ 
   \hline 
   \multirow{6}{*}{0.2}
   &GMDI & 	\textbf{1.67}(0.53) & 	\textbf{0.3}(0.12) \\ 
  & TBMDE & 1.78(0.68) & 0.32(0.2) \\ 
  & mice & 3.1(1.53) & 1.58(1.7) \\ 
  & Hmisc & 2.42(1.19) & 0.88(0.85) \\ 
  & Amelia & 3.04(1.22) & 5.05(1.35) \\ 
  & mi & 2.82(1.02) & 5.47(1.5) \\ 
   \hline
    \multirow{6}{*}{0.4}
       &GMDI & 	\textbf{2.28}(0.93) & 	\textbf{0.79}(0.41) \\ 
  &TBMDE & 2.43(0.81) & 0.82(0.38) \\ 
  &mice & 4.58(1.8) & 2.8(1.99) \\ 
  &Hmisc & 3.12(1.21) & 2.61(1.62) \\ 
  &Amelia & 2.8(0.75) & 5.67(2.75) \\ 
  &mi & 3.13(0.76) & 6.09(2.83) \\ 
  \hline

\end{tabular}
    \caption{MSE of true data and prediction in GMDI, TBMDE and the other four R packages for the simulation data ($n=100$). The bold values are the best ones in each column.}
    \label{tab:sim_data}
\end{table}

\begin{figure}
\centering
\includegraphics[width=1\textwidth]{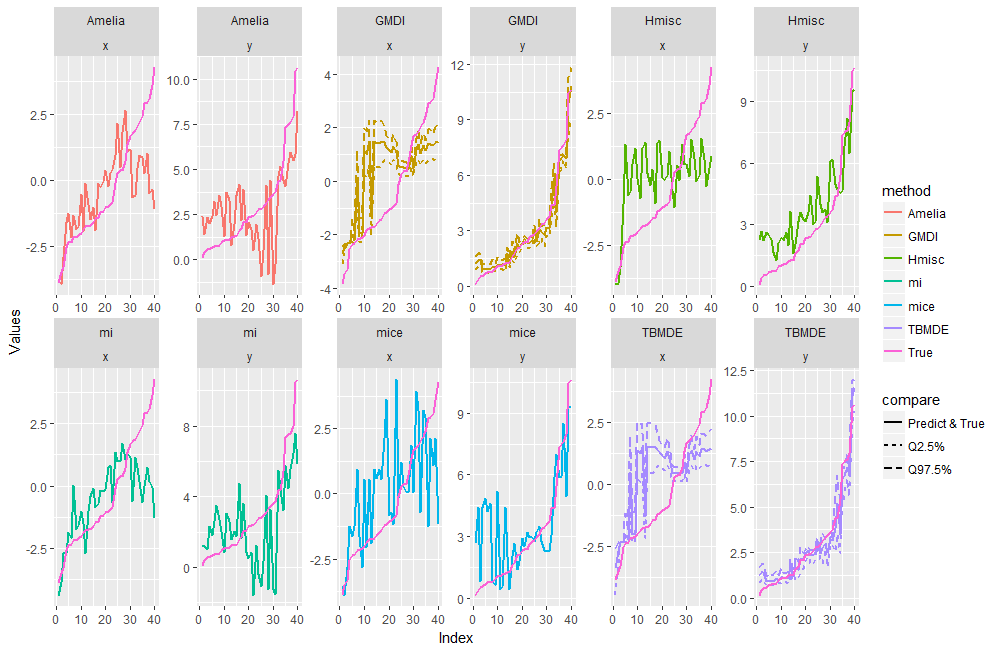}
\caption{\label{fig:PRED1}Predition of missing values of $X$ and $Y$ for the simulation data when $n=100$, $r=0.4$. Pink lines correspond to the true values of missing data. Solid lines correspond to the mean predicted values of missing data while dashed lines correspond to the 2.5\% and 97.5\% quantiles of the predicted values with respect to sampled $\boldsymbol{(\theta, \lambda)}$'s. Six methods of missing data imputation are compared.}
\end{figure}

\begin{table}[]
    \centering
    \begin{tabular}{c|c|cccc}
  \hline
  \hline
 r & Methods & $x_1$ & $x_2$ & $x_3$ & $x_4$ \\ 
  \hline
  \hline
 	\multirow{7}{*}{0.1}&
GMDI & 	\textbf{361.24}(190.32) & 9946.87(3586.44) & 11.83(5.92) & 48.08(21.81) \\ 
 & TBMDE & 370.29(195.99) & 10141.28(3628.4) & 11.95(5.69) & 48.37(22.21) \\ 
&  mice & 804.65(565.99) & 14749.9(5639.25) & 19.8(10.21) & 86.56(38.28) \\ 
 & Hmisc & 500.27(430.11) & 	\textbf{9143.69}(3725.81) & 12.25(6.66) & 50.64(16.66) \\ 
  &missForest & 414.2(428.79) & 10351.34(4001.89) & 	\textbf{11.37}(6.75) & 	\textbf{43.53}(18.62) \\ 
  &Amelia & 606.19(391.57) & 9459.01(3518.68) & 11.51(6.58) & 59.07(24.31) \\ 
  &mi & 679.03(386.11) & 11152.91(3363.03) & 11.7(5.29) & 62.92(18.68) \\ 
   \hline 
   \multirow{7}{*}{0.2}
   &GMDI & 476.72(306.58) & 8749.21(2503.08) & 11.42(4.17) & 	\textbf{54.57}(16.99) \\ 
  & TBMDE & 495.66(309.63) & 9477.48(2789.18) & 11.73(4.26) & 57.26(17.07) \\ 
  &mice & 865.65(403.69) & 14875.61(5045.66) & 14.58(4.01) & 98.39(33.06) \\ 
  &Hmisc & 578.25(341.67) & 9121.95(2593.14) & 11.89(3.41) & 65.06(17.03) \\ 
  &missForest & 	\textbf{446.44}(275.08) & 	\textbf{8700.26}(2195.47) & 11.54(3.74) & 56.38(18.78) \\ 
  &Amelia & 617.94(256.88) & 9335.43(1976.33) & 	\textbf{10.75}(3.45) & 63.74(16.65) \\ 
  &mi & 682.18(283.73) & 9603.42(1756.92) & 12.36(3.92) & 71.75(18.91) \\ 
   \hline
    \multirow{7}{*}{0.4}
    &GMDI& \textbf{593.52}(186.42) & 9815.57(2033.09) & 11.01(2.7) & 	\textbf{58.33}(13.11) \\ 
  & TBMDE & 665.91(207.32) & 11361.53(2688.63) & 12.58(3.57) & 64.96(16.24) \\ 
  &mice & 1128.2(320.36) & 15418.4(3292.92) & 18.37(3.82) & 104.67(23.83) \\ 
  &Hmisc & 731.23(183.4) & 	\textbf{9401.04}(1833.13) & 11.54(2.09) & 65.69(13.14) \\ 
  &missForest & 659.4(223.11) & 9782.48(2265.9) & 11.73(2.47) & 62.95(10.17) \\ 
  &Amelia & 757.39(191.72) & 9754.93(1716.32) & 	\textbf{10.61}(2.55) & 65.35(14.3) \\ 
 & mi & 805.02(203.63) & 10054.78(1649.03) & 11.44(2.28) & 71.73(15.78) \\ 

   \hline
\end{tabular}
    \caption{MSE of true data and prediction in GMDI, TBMDE and the other five R packages for the real dataset (airquality). The bold values are the best ones in each column.}
    \label{tab:airquality}
\end{table}

\begin{figure}
\centering
\includegraphics[width=1\textwidth]{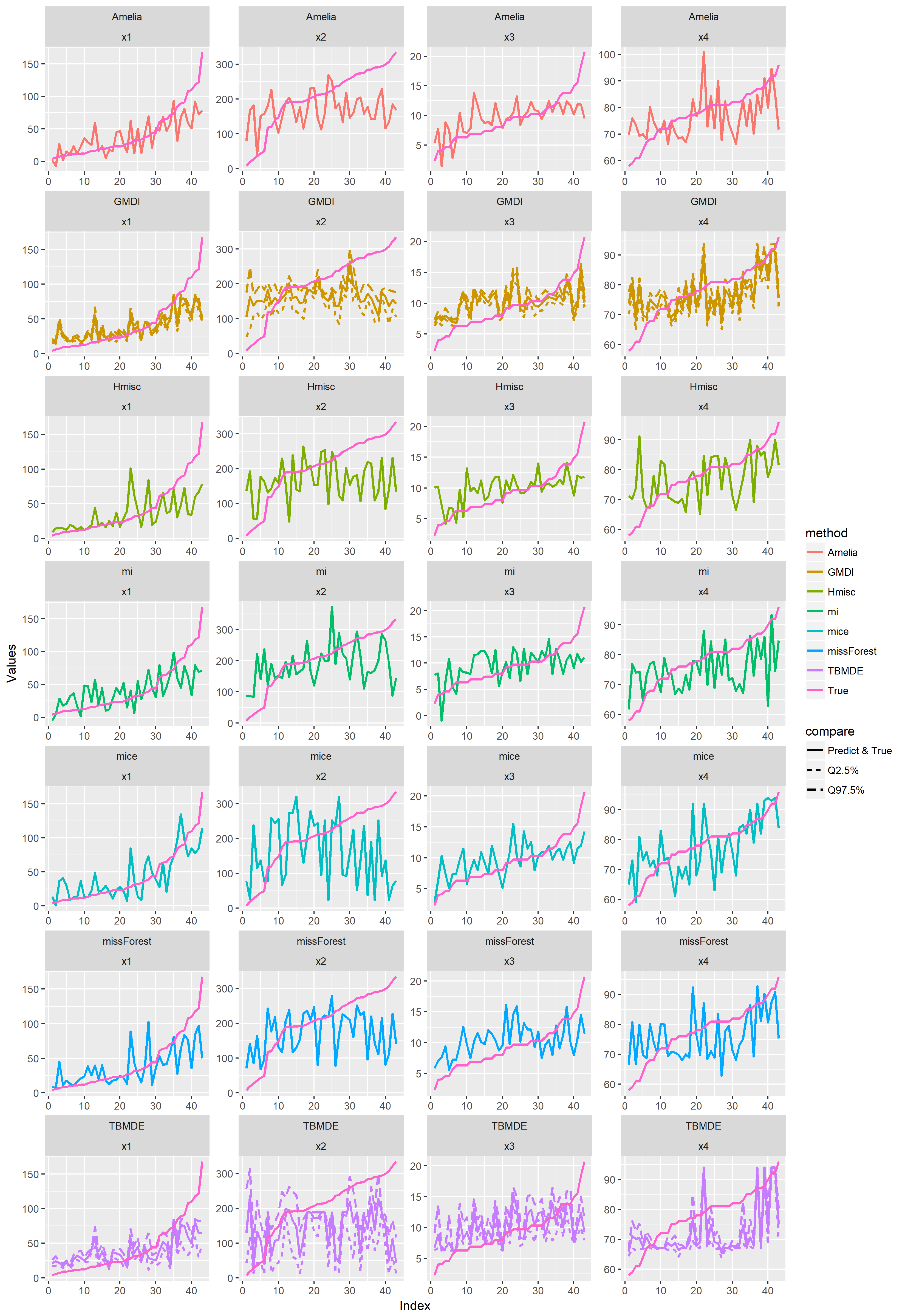}
\caption{\label{fig:PRED2}Predition of missing values of $X_1$, $X_2$, $X_3$ and $X_4$ for the ``airquality' data when $n=111$, $r=0.4$. Pink lines correspond to the true values of missing data. Solid lines correspond to the mean predicted values of missing data while dashed lines correspond to the 2.5\% and 97.5\% quantiles of the predicted values with respect to sampled $\boldsymbol{(\theta,\lambda)}$'s. Seven methods of missing data imputation are compared.}
\end{figure}

\begin{figure}
\centering
\includegraphics[width=1\textwidth]{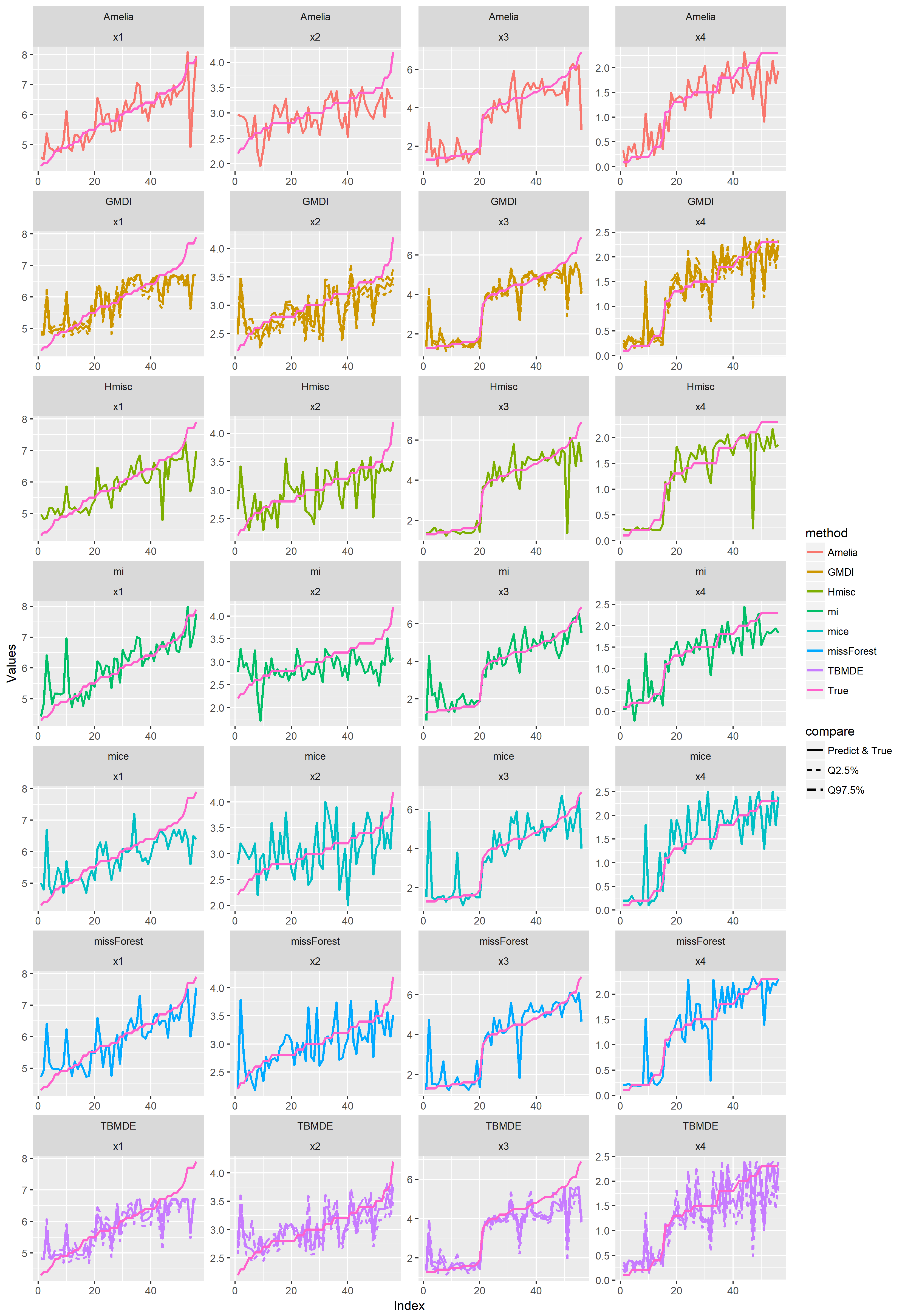}
\caption{\label{fig:PRED3}Predition of missing values of $X_1$, $X_2$, $X_3$ and $X_4$ for the ``Iris'' data when $n=150$, $r=0.4$. Pink lines correspond to true values of missing data. Solid lines correspond to the mean predicted values of missing data while dashed lines correspond to the 2.5\% and 97.5\% quantiles of the predicted values with respect to sampled $\boldsymbol{(\theta, \lambda)}$'s. Seven methods of missing data imputation are compared.}
\end{figure}

\begin{table}[]
    \centering
    \begin{tabular}{c|c|cc}
    \hline
    \hline
   r & Variable & KSP(GMDI) & KSP(TBMDE)  \\
    \hline
    \hline
    \multirow{4}{*}{0.1}
   & $x_1$ & 	\textbf{0.29}(0.17) & 0.26(0.17) \\ 
  &$x_2$ & 	\textbf{0.60}(0.28) & 0.45(0.28) \\ 
  &$x_3$ & 	\textbf{0.32}(0.23) & 0.27(0.24) \\ 
  &$x_4$ & 	\textbf{0.56}(0.27) & 0.51(0.28) \\ 
    \hline
    \multirow{4}{*}{0.2}
      & $x_1$ &  	\textbf{0.28}(0.19) & 0.24(0.17) \\ 
   &$x_2$  & 	\textbf{0.51}(0.32) & 0.25(0.3) \\ 
  &$x_3$& 	\textbf{0.31}(0.2) & 0.17(0.18) \\ 
  &$x_4$ & \textbf{0.48}(0.25) & 0.2(0.16) \\ 
    \hline
    \multirow{4}{*}{0.4}& $x_1$ & \textbf{0.10}(0.14) &0.04(0.19)\\ 
    
    & $x_2$ & \textbf{0.16}(0.24) & 0.06(0.14)\\
    
    & $x_3$ & \textbf{0.12}(0.18) & 0.04(0.06)\\
    
    & $x_4$ & \textbf{0.16}(0.22) &0.07(0.09)\\
    \hline

    \end{tabular}
    \caption{KS test p-values of density estimation by GMDI and TBMDE for the real dataset (airquality). The bold values are the best ones in each row.}
    \label{tab:airquality_ks}
\end{table}
\end{document}